\documentclass[10pt]{article}
\usepackage{amssymb}
\usepackage{stmaryrd}
\title{Introduction to linear logic and ludics, Part II}
\author{Pierre-Louis Curien (CNRS \& Universit\'e Paris VII)\footnote{Laboratoire {\it Preuves, Programmes et Syst\`emes}, Case 7014, 2 place Jussieu, 75251 Paris Cedex 05, France, curien@pps.jussieu.fr.}}

\begin{document}

\maketitle

\newtheorem{theorem}{Theorem}[section]
\newtheorem{definition}[theorem]{Definition}
\newtheorem{lemma}[theorem]{Lemma}
\newtheorem{corollary}[theorem]{Corollary}
\newtheorem{proposition}[theorem]{Proposition}
\newtheorem{example}[theorem]{Example}
\newtheorem{exercise}[theorem]{Exercise}
\newtheorem{hardexercise}[theorem]{Exercise${}^*$}
\newtheorem{notation}[theorem]{Notation}
\newtheorem{remark}[theorem]{Remarká}

\newenvironment{branch}{\left\{\begin{array}{l}}{\end{array}\right.}

\newcommand{\qed}{\hfill$\Box$}
\newcommand{\qedm}{\mbox{}\\[-2.5em] \mbox{}\hfill $\Box$} 
\newcommand{\Proof}{\noindent {\sc Proof}. }
\newcommand{\Proofhint}{\noindent {\sc Proof (indication)}. }
\newcommand{\Proofitem}[1]{\medskip \noindent $#1\;$}
\newcommand{\Proofitemf}[1]{ $#1\;$}

\newcommand{\guil}[1]{``#1"}
\newcommand{\lbd}{\lambda}
\newcommand{\mand}{\mbox{ and }}
\newcommand{\mor}{\mbox{ or }}
\newcommand{\nin}{\not\in}
\newcommand{\inc}{\subseteq}
\newcommand{\set}[1]{\{#1\}}
\newcommand{\setc}[2]{\set{#1 \mid #2}}
\newcommand{\faux}{{\bf F}}
\newcommand{\et}{\wedge}
\newcommand{\ou}{\vee}
\newcommand{\vide}{\emptyset}
\newcommand{\implies}{\Rightarrow}
\newcommand{\si}{\Leftarrow}
\renewcommand{\iff}{\Leftrightarrow}
\newcommand{\qqs}[2]{\forall\, #1\;\: #2}
\newcommand{\xst}[2]{\exists\, #1\;\: #2}
\newcommand{\nxst}[2]{\not\!\exists\, #1\;\: #2}
\newcommand \seql[3]{\raisebox{3ex}{$\mbox{#1}\;\;$} \; \shortstack{$#2$ \\ \mbox{}\\
                    \mbox{}\hrulefill\mbox{}\\ \mbox{}\\ $#3$}}
\newcommand \seq[2]{\shortstack{$#1$ \\ \mbox{}\\
                    \mbox{}\hrulefill\mbox{}\\ \mbox{}\\ $#2$}}

\newcommand{\forces}{\makebox[5mm]{\,$\|\!-$}}%

\newcommand{\fun}{\:\rightarrow\:} 
\newcommand{\funt}{\rightarrow^{\ast}} 
\newcommand{\lfun}[1]{\stackrel{#1}{\longrightarrow}}
\newcommand{\lfunt}[1]{\stackrel{#1}{\longrightarrow}\,\!^{\ast}}
\newcommand{\length}[1]{ | #1 |}
\newcommand{\union}{\cup}		
\newcommand{\inter}{\cap}		
\newcommand{\Union}{\bigcup}		
\newcommand{\Inter}{\bigcap}		
\newcommand{\join}{\vee}		
\newcommand{\JOIN}{\bigvee}		
\newcommand{\meet}{\wedge}		
\newcommand{\MEET}{\bigwedge}		

\newcommand{\inv}[1]{#1^{-1}}
\newcommand{\card}{\sharp}		
\newcommand{\pfun}{\rightharpoonup}  
\newcommand{\ul}[1]{\underline{#1}}	

\newcommand{\id}{{\it id}}
\newcommand{\catch}{{\it catch}\;}
\newcommand{\throw}{{\it throw}\;}
\newcommand{\Alt}{ \mid\!\!\mid  } 
\newcommand{\Sub}[3]{#1[#2\leftarrow #3]}	
\newcommand{\sub}[2]{[#2 / #1]}
\newcommand{\subn}[2]{[#2 /_{\!\circ} #1]}
\newcommand{\BO}[4]{#1\,,#2\stackrel{#3}{\Longrightarrow} #4}
\newcommand{\bt}[1]{{\it BT}(#1)}
\newcommand{\lelong}[2]{#1\!\mid_{#2}}	
\newcommand{\dl}{[\![} 			
\newcommand{\dr}{]\!]} 			
\newcommand{\newv}{{\it new}\:}
\newcommand{\deref}{{\it deref}\:}
\newcommand{\nat}{{\bf int}}
\newcommand{\varimp}{{\bf var}\:}
\newcommand{\comm}{{\bf comm}}
\newcommand{\world}{{\bf W}}
\newcommand{\comp}{\circ}
\newcommand{\run}{{\it run}}
\newcommand{\done}{{\it done}}
\newcommand{\readv}{{\it read}}
\newcommand{\writev}[1]{{\it write}(#1)}
\newcommand{\OK}{{\it OK}}
\newcommand{\lpar}{\bindnasrepma}
\newcommand{\with}{\&}
\newcommand{\limpl}{\multimap}
\newcommand \coh {\mathrel{\raisebox{-.3em}{\shortstack{$\frown$\\$\smile$}}}}
\newcommand \icoh {\mathrel{\raisebox{-.3em}{\shortstack{$\smile$\\$\frown$}}}}
\newcommand \seqdbl[2]{\shortstack{$#1$ \\ \mbox{}\\
                    \mbox{}\hrulefill\mbox{}\\  \mbox{}\hrulefill\mbox{}\\ \mbox{}\\ $#2$}}

\newcommand{\bbot}{\bot}
\newcommand{\rst}[1]{\!\upharpoonright^{#1}} 
\newcommand{\cat}[1]{{\bf #1}}
\newcommand{\kleisli}[1]{\kappa(#1)}
\newcommand{\pair}[2]{\langle #1 , #2 \rangle}
\newcommand{\prfnet}[2]{\{#1\}[#2]}
\newcommand{\prfnetc}[3]{[#1]\{#2\}[#3]}

\newcommand{\demon}{\maltese}
\newcommand{\reseau}[2]{\langle #1 \:|\: #2 \rangle}
\newcommand{\ppm}[2]{[#1,\stackrel{#2}{\hookleftarrow}]}
\newcommand{\sqinc}{\sqsubseteq}
\newcommand{\coupe}[2]{\langle #1 \:|\: #2 \rangle}
\newcommand{\dirct}[1]{{\it Dir}(#1)}
\newcommand{\tensorr}{<\!\!\!\!\!\!\!\bigcirc}
\newcommand{\makep}[1]{\downarrow\! #1}
\newcommand{\maken}[1]{\uparrow\! #1}
\newcommand{\stoup}[2]{\vdash #1 ; #2}
\newcommand{\prodcom}{\Union\!\!\!\!\!\!\times\:}

\begin{abstract}
This paper is the second part of an introduction to  linear logic and ludics, both due to Girard. 
It is devoted to proof nets, in the limited, yet central, framework of multiplicative linear logic (section \ref{proof-nets}) and to ludics, which has been recently developped in an aim of further unveiling the fundamental interactive nature of computation and logic (sections \ref{ludics}, \ref{designs}, \ref{normalization}, and \ref{behaviours}). We hope to offer a few computer science insights into this new theory. 
\end{abstract}


\subsection*{Prerequisites}  This part depends mostly on the first two sections  of part I, and should therefore be accessible to any one who has been exposed to the very first steps of linear logic. 
We have used the following sources:  \cite{Gir87,DR89, GueMa2000} for proof nets, and  and \cite{LS00,Gir01} for ludics.

\section{Proof nets} \label{proof-nets}
Proof nets are graphs that give a more economical presentation of proofs, abstracting from some irrelevant order of application of the rules of linear logic given in sequent calculus style (sections
2 and 3 of part I). We limit ourselves here to multiplicative linear logic (MLL), and we shall even begin with cut-free MLL.  
Proof nets for MLL are graphs which are ``almost trees'', and we stress this by our choice of presentation. 
We invite the reader to draw proof nets by himself while reading the section.

What plays here the role of a proof of a sequent $\vdash A_1,\ldots,A_n$ is
the forest of the formulas $A_1,\ldots,A_n$ represented as trees, together with a partition
of the leaves of the forest in pairs (corresponding to the application of an axiom $\vdash C,C^\bot$).  Well, not quite. A proof may not decompose each formula completely, since
an axiom $\vdash C,C^\bot$ can be applied to {\em any} formula. Hence we have to specify
{\em partial} trees for the formulas $A_1,\ldots, A_n$. One way to do this is to specify the set of leaves of each partial tree as a set of (pairwise disjoint) occurrences.  Formally, an occurrence is a word over $\set{1,2}$, 
and the subformula $A/u$ at occurrence $u$ is defined by the following formal system:
$$A/\epsilon = A\quad\quad (A_1\otimes A_2)/1u=(A_1\lpar A_2)/1u=A_1/u
\quad\quad (A_1\otimes A_2)/2u=(A_1\lpar A_2)/2u=A_2/u\;.$$

\noindent 
A {\em partial formula tree} $A^U$ consists of a formula $A$ together with
a set $U$ of pairwise disjoint occurrences  such that $(\qqs{u\in U}{A/u\mbox{ is defined}})$.
Recall that a partition of a set $E$ is a set $X$ of non-empty subsets of $E$ which are pairwise disjoint and whose union is $E$.  If $X=\set{Y_1,\ldots,Y_n}$ we often simply write $X=Y_1,\ldots,Y_n$.

\begin{definition}
A {\em proof structure} is given by 
$$\prfnet{A_1^{U_1},\ldots,A_n^{U_n}}{X}\;,$$ where $\set{A_1^{U_1},\ldots,A_n^{U_n}}$ is a multiset of partial formula trees and
where $X$ is a partition of  $\setc{A_1/u}{u\in U_1}\union\ldots\union\setc{A_n/u}{u\in U_n}$ (disjoint union) whose classes are pairs of dual formulas.
We shall say that each class of the partition is an axiom  of the proof structure, and that 
$A_1,\ldots,A_n$ are the conclusions of the proof structure.
\end{definition}

More generally, we shall manipulate graphs described as a forest plus a partition of its leaves, without any particular requirement on the partition. This notion is faithful to Girard's idea of paraproof, discussed in the next section. 

\begin{definition} \label{paraproof-structure}
A {\em paraproof structure} is given by $\prfnet{A_1^{U_1},\ldots,A_n^{U_n}}{X}$, where $X$ is a partition of  $\setc{A_1/u}{u\in U_1}\union\ldots\union\setc{A_n/u}{u\in U_n}$ (disjoint union) .
We shall say that each class of the partition is a generalized axiom , or {\em daimon} (anticipating on a terminology introduced in the following section), and that 
$A_1,\ldots,A_n$ are the conclusions of the paraproof structure.
\end{definition}

The actual graph associated to this description is obtained by:
\begin{itemize}
\item drawing the trees of the formulas $A_1,\ldots,A_n$ stopping at $U_1,\ldots,U_n$, respectively (i.e., there is a node corresponding to each subformula $A_i/u$, where $u<u_i$ for some $u_i\in U_i$); and
\item associating a new node with each class  of the partition and
new edges from the new node to each of the leaves of the class (in the case of proof structures, one can dispense with the new node, and draw an edge between the matching dual formulas -- such edges are
usually drawn horizontally).
\end{itemize}

We next describe how to associate a proof structure with a proof. The following definition
follows the rules of MLL.

\begin{definition} \label{NET-criterion}
The {\em sequentializable} proof structures for (cut-free) MLL are the proof structures obtained by the following rules:

$$\begin{array}{c}
\seq{}{\prfnet{C^{\set{\epsilon}},(C^\bot)^{\set{\epsilon}}}{\set{\set{C,C^\bot}}}} \quad\quad
\seq{\prfnet{A_1^{U_1},\ldots,A_n^{U_n},B^V,C^W}{X}}{\prfnet{A_1^{U_1},\ldots,A_n^{U_n},(B\lpar C)^U}{X}}\\\\\
\seq{\prfnet{A_1^{U_1},\ldots,A_n^{U_n},B^V}{X}\quad\prfnet{{A'}_1^{U'_1},\ldots,{A'}_{n'}^{U'_{n'}},C^W}{Y}}{\prfnet{A_1^{U_1},\ldots,A_n^{U_n},{A'}_1^{U'_1},\ldots,{A'}_{n'}^{U'_{n'}},(B\otimes C)^U}{X\union Y}}
\end{array}$$

\noindent
where $U=\setc{1v}{v\in V}\union\setc{2w}{w\in W}$.  Sequentializable proof structures are called {\em  proof nets}.
\end{definition}

It should be clear that there is a bijective correspondence between MLL proofs and the 
proofs  of sequentialization. Let us check one direction. We show that if a proof structure with conclusions $A_1,\ldots, A_n$  is sequentializable, then a proof that it is so yields an MLL proof of the sequent $\vdash A_1,\ldots,A_n$. This is easily seen by induction: the proof associated with $\prfnet{C^{\set{\epsilon}},(C^\bot)^{\set{\epsilon}}}{\set{\set{C,C^\bot}}}$ is the axiom $\vdash C,C^\bot$, while the last steps of the proofs associated with 
$\prfnet{A_1^{U_1},\ldots,A_n^{U_n},(B\lpar C)^U}{X}$ and 
$\prfnet{A_1^{U_1},\ldots,A_n^{U_n},{A'}_1^{U'_1},\ldots,{A'}_n^{U'_n},(B\otimes C)^U}{X}$ are, respectively:

$$\seq{\vdash A_1,\ldots,A_n,B,C}{\vdash A_1,\ldots,A_n,B\lpar C}\quad\quad
\seq{\vdash A_1,\ldots,A_n,B\quad\vdash A'_1,\ldots A'_{n'},C}{\vdash A_1,\ldots,A_n,A'_1,\ldots,A'_{n'},B\otimes C}$$

\smallskip
The question we shall address in the rest of the section is the following: given a  proof structure, when is it the case that it is sequentializable?  It turns out that the right level of generality for this question is to lift it to  paraproof nets, which are defined next.

\begin{definition} The {\em sequentializable} paraproof structures for (cut-free) MLL are the paraproof structures obtained by the following rules:

$$
\seq{}{\prfnet{A_1^{\set{\epsilon}},\ldots,A_n^{\set{\epsilon}}}{\set{\set{A_1,\ldots,A_n}}}}\quad\quad
\seq{\prfnet{A_1^{U_1},\ldots,A_n^{U_n},B^V,C^W}{X}}{\prfnet{A_1^{U_1},\ldots,A_n^{U_n},(B\lpar C)^U}{X}}$$

$$\seq{\prfnet{A_1^{U_1},\ldots,A_n^{U_n},B^V}{X}\quad
\prfnet{{A'}_1^{U'_1},\ldots,{A'}_{n'}^{U'_{n'}},C^W}{Y}}{\prfnet{A_1^{U_1},\ldots,A_n^{U_n},{A'}_1^{U'_1},\ldots,{A'}_{n'}^{U'_{n'}},(B\otimes C)^U}{X\union Y}}$$

\noindent
where $U=\setc{1v}{v\in V}\union\setc{2w}{w\in W}$. Sequentializable paraproof structures are called {\em  paraproof nets}. 
\end{definition}

What is the proof-theoretic meaning of a paraproof net? Well, the above bijective correspondence 
extends to a correspondence between sequentialization proofs of paraproof structures and
MLL {\em paraproofs}, which are defined by the following rules:

$$\seq{}{\vdash\Gamma}\quad\quad\quad\seq{\vdash\Gamma_1,B\quad\vdash\Gamma_2,C}{\vdash\Gamma_1,\Gamma_2,B\otimes C}\quad\quad\quad\seq{\vdash\Gamma,B,C}{\vdash\Gamma,B\lpar C}$$

\noindent 
where in the first rule $\Gamma$ is an arbitrary multiset of formulas. The first rule is called {\em generalized axiom}, or {\em daimon} rule.
Starting in a proof search mode from an MLL formula $A$ one may build absolutely freely a paraproof of $A$, making arbitrary decisions when splitting the context in a $\otimes$ rule. Any choice is as good as another, in the sense that the process will be successful at the end, i.e., we shall eventually reach generalized axioms. 
There is an interesting subset of paraproofs, which we call  the {\em extreme} ones. An extreme paraproof is a paraproof in which in each application of the $\otimes$ rule we have $\Gamma_1=\emptyset$ or $\Gamma_2=\emptyset$. The following definition formalizes this notion.

\begin{definition} The {\em extreme} paraproof nets for (cut-free) MLL are the proof structures obtained by the following rules:

$$\begin{array}{c}
\seq{}{\prfnet{A_1^{\set{\epsilon}},\ldots,A_n^{\set{\epsilon}}}{\set{\set{A_1,\ldots,A_n}}}}\quad\quad
\seq{\prfnet{A_1^{U_1},\ldots,A_n^{U_n},B^V,C^W}{X}}{\prfnet{A_1^{U_1},\ldots,A_n^{U_n},(B\lpar C)^U}{X}}\\\\
\seq{\prfnet{A_1^{U_1},\ldots,A_n^{U_n},B^V}{X}\quad
\prfnet{C^W}{Y}}{\prfnet{A_1^{U_1},\ldots,A_n^{U_n},(B\otimes C)^U}{X\union Y}}
\quad\quad
\seq{\prfnet{B^V}{X}\quad
\prfnet{{A}_1^{U_1},\ldots,{A}_{n}^{U_{n}},C^W}{Y}}{\prfnet{A_1^{U_1},\ldots,A_n^{U_n},(B\otimes C)^U}{X\union Y}}
\end{array}$$

\noindent
where $U=\setc{1v}{v\in V}\union\setc{2w}{w\in W}$.
\end{definition}

Extreme paraproofs will be soon put to use, but for the time being we are concerned with general paraproofs.
Our first tool is the notion of switching. Let $S$ be a paraproof stucture. A switching is
a function from the set of all internal nodes of (the forest part of) $S$ of the form
$B_1\lpar B_2$, to $\set{L,R}$.  A switching induces a {\em correction} (sub)graph, defined as follows.
At each internal node $B_1\lpar B_2$, we cut exactly one of the two edges linking  $B_1\lpar B_2$ to its immediate subformulas, namely the edge between $B_1\lpar B_2$ and $B_2$ if the switching is set to $L$, and the edge between $B_1\lpar B_2$ and $B_1$ if the switching is set to $R$.
The following definition is due to Danos and Regnier \cite{DR89}, and arose as a simplification of the original criterion proposed by Girard \cite{Gir87}.

\begin{definition}[DR] We say that a paraproof stucture $S$ satisfies the {\em DR-criterion} if all the correction graphs induced by a switching of $S$ are connected and acyclic (that is, are trees).
\end{definition}

\begin{proposition} \label{strong-DR}
All sequentializable paraproof structures satisfy the DR-criterion.
Moreover, in each correction graph, when a switch corresponding to a formula $B_1\lpar B _2$ is, say, on the left, then the path from $B_1$ to $B_2$ does not go through $B_1\lpar B_2$.
\end{proposition}

\Proof
We proceed by induction on the definition of paraproof nets. If $N$ is a generalized axiom,
then there is just one switching (the empty one)  and the associated correction graph is the graph 
itself, which is obviously a tree. If $N$ is obtained from $N_1$ and $N_2$ by a $\otimes$ rule 
acting on a conclusion $B$ of $N_1$ and a conclusion $C$ of  $N_2$, then a switching of $N$ is a pair of a switching of $N_1$ and a switching of $N_2$. We know by induction that the corresponding two correction graphs are trees. We can thus organize them in a tree form, with roots $B$, $C$ respectively. Then the correction graph for $N$  is obtained by adding a new root and edges between the new root and $B$ and $C$, respectively: this is obviously a tree.
Finally, suppose that  $N$ is obtained from $N_1$ by a $\lpar$ rule acting on two conclusions $B$ and $C$ of $N_1$, and consider a  switching for $N$, which assigns, say $L$, to the new $\lpar$ node. The rest of the switching determines a correction graph for $N_1$ which is a tree by induction. We can 
organize the correction graph for $N_1$  in such a way that $B$ is the root. Then the correction graph for $N$ is obtained by adding a new root and an edge between the new root and $B$, and this is again obviously a tree.
The second property of the statement is obvious to check, by induction on the sequentialization  proof too. \qed

\medskip
Our next tool is a parsing procedure, which takes a paraproof structure and progressively shrinks it, or contracts it. If the procedure is not blocked, then the paraproof structure is a paraproof net.
This criterion was first discovered by Danos \cite{DanosTh}. Guerrini explained the criterion as a successful parsing procedure \cite{Gue99}. Here we (straightforwardly) extend the procedure from proof structures to paraproof structures.

\begin{definition} \label{PARSING-criterion}
We define the following rewriting system on paraproof structures:

$$\begin{array}{c}
\seq{A/u=B_1\lpar B_2}{{\prfnet{\Gamma,A^{U\union\set{u1,u2}}}{X,\set{\Delta,B_1,B_2}}\rightarrow_P
\prfnet{\Gamma,A^{U\union\set{u}}}{X,\set{\Delta,(B_1\lpar B_2)}}}} \\\\
\seq{A/u=B_1\otimes B_2}{{\prfnet{\Gamma,A^{U\union\set{u1,u2}}}{X,\set{\Delta_1,B_1},\set{\Delta_2,B_2}} \rightarrow_P
\prfnet{\Gamma,A^{U\union\set{u}}}{X,\set{\Delta_1,\Delta_2,B_1\otimes B_2}}}}
\end{array}$$

\noindent We say that a proof structure $S=\prfnet{A_1^{U_1},\ldots,A_n^{U_n}}{X}$ satisfies the {\em weak Parsing criterion} if $$S\rightarrow_P^\star \prfnet{A_1^{\set{\epsilon}},\ldots,A_n^{\set{\epsilon}}}{\set{A_1,\ldots,A_n}}$$
and that it satisfies the {\em strong Parsing criterion} if  any reduction sequence $S\rightarrow_P^\star S'$ can be completed by a reduction $S'\rightarrow_P^\star \prfnet{A_1^{\set{\epsilon}},\ldots,A_n^{\set{\epsilon}}}{\set{A_1,\ldots,A_n}}$.
\end{definition}

\begin{lemma} \label{DR-stable-PARSING}
 If $S\rightarrow_P S'$, and if $S$ satisfies the DR-criterion, then $S'$ satisfies the DR-criterion.
\end{lemma}

\Proof Suppose that $S\rightarrow_P S'$ by the $\lpar$ rule, and that a switching for $S'$ has been fixed. $S$ is the same graph as $S'$ except that $S$ has two additional vertices $B_1$ and $B_2$ and that the edge connecting $B_1\lpar B_2$ with its class in $S'$ is replaced by
a diamond of edges between $B_1\lpar B_2$, $B_1$, $B_2$, and its class $\set{\Delta,B_1,B_2}$ in $S$.
We extend the switching to $S$ by assigning, say $L$, to $B_1\lpar B_2$ (which is internal in $S$). By assumption, the correction graph is a tree. We can take $B_2$ as a root, which has the class $\set{\Delta,B_1,B_2}$ as unique son, which has $B_1$ among his sons, which has $B_1\lpar B_2$ as unique son. Then the correction graph for the original switching relative to $S'$ is obtained by collapsing $B_2$ with $\set{\Delta,B_1,B_2}$ and $B_1$ with $B_1\lpar B_2$, and is still a tree.

Suppose now that  $S\rightarrow_P S'$ by the $\otimes$ rule. A switching of $S'$ is also a switching for $S$, whose associated correction graph is thus a tree. Let us take $B_1\otimes B_2$ as a root.
Then  the correction graph for $S'$ is obtained as follows: collapse $B_1$, its unique son $\set{\Delta_1,B_1}$,  $B_2$, and its unique son $\set{\Delta_2,B_2}$.
This clearly yields a tree.  \qed

\begin{proposition} If a proof structure satisfies the DR-criterion, then it satisfies the strong Parsing criterion.
\end{proposition}

\Proof Let $S$ be a paraproof structure with conclusions $A_1,\ldots,A_n$.
Clearly, each $\rightarrow_P$ reduction strictly decreases the size of the underlying forest, hence all reductions terminate.
Let $S\rightarrow_P^{\star} S'$, where $S'$ cannot be further reduced. We know by Lemma 
\ref{DR-stable-PARSING} that $S'$ also satisfies the DR-criterion.
We show that $S'$ must be a generalized axiom (which actually entails that $S'$ must precisely be $\prfnet{A_1^{\set{\epsilon}},\ldots,A_n^{\set{\epsilon}}}{\set{A_1,\ldots,A_n}}$, since the set of conclusions remains invariant under reduction). Suppose that $S'$ is not a generalized axiom, and consider an arbitrary class $\Gamma$ of the partition of $S'$. We claim that $\Gamma$ contains at least one formula whose father is a $\otimes$ node. Indeed, otherwise, each element of the class is either a conclusion or a formula $B$ whose father is a $\lpar$ node. Note that in the latter case the other child of the $\lpar$ node cannot belong to the class as otherwise $S'$ would not be in normal form. If the father of $B$ is $B\lpar C$ (resp. $C\lpar B$), we set its switch to $R$ (resp. $L$), and we extend the switching arbitrarily to all the other internal $\lpar$ nodes of $S'$.
Then the restriction of the correction graph $G$  to $\Gamma$ and its elements forms a connected component, which is strictly included in $G$ since $S'$ is not a generalized axiom. But then $G$ is not connected, which is a contradiction.

We now construct a path in $S'$ as follows. We start from a leaf $B_1$ whose father is a $\otimes$ node, say $B_1\otimes C$, we go down to its father and then up  through $C$ to a leaf $B'_1$, choosing a path of maximal length. All along the way, when we meet $\lpar$ nodes, we set the switches in such a way that the path will remain in the correction graph.
$B'_1$ cannot have a $\otimes$ node as father, as otherwise by maximality the other son $C'_1$ of this father would be a leaf too, that cannot belong to the same class as this would make a cycle, and cannot belong to another class because $S'$ is in normal form.
Hence, by the claim, we can pick $B_2$ different from $B'_1$ in its class whose father is a $\otimes$ node. We continue our path by going up from $B'_1$ to its class, and then down to $B_2$, down to its father, and we consider again a maximal path upwards from there. Then this path cannot meet any previously met vertice, as otherwise, setting the switches in the same way as above until this happens, we would get a cycle in a correction graph. Hence we can reach a leaf $B'_2$ without forming a cycle, and we can continue like this forever. But $S'$ is finite, which gives a contradiction. \qed

\begin{proposition} If a paraproof structure satisfies the weak Parsing criterion, then it is sequentiali\-zable.
\end{proposition} 

\Proof
We claim that if $S\rightarrow_P^\star S'$, then $S$ can be obtained from $S'$ by replacing each generalized axiom of $S'$ by an appropriate paraproof net. We proceed by induction on the length of the derivation. In the base case, we have $S'=S$, and we replace each generalized axiom by itself. Suppose that
$S\rightarrow_P^\star S'_1\rightarrow_P S'$. We use the notation of Definition \ref{PARSING-criterion}. Suppose that $S'_1\rightarrow_P S'$ by the $\lpar$ reduction rule. By induction, we have a paraproof net $N$ to substitute for $\set{\Delta,B_1,B_2}$. We can add a $\lpar$ node to $N$ and get a new paraproof net  which when substituted for $\set{\Delta,B_1\lpar B_2}$ in $S'$ achieves the same effect as the substitution of $N$ in $S'_1$, keeping the same assignment of
paraproof nets for all the other generalized axioms.
 The case of the $\otimes$ rule is similar: we now have by induction two paraproof nets $N_1$ and $N_2$ to substitute for 
$\set{\Delta_1,B_1}$ and $\set{\Delta_2,B_2}$, and we form a new paraproof net by the $\otimes$ rule which when substituted  for $\set{\Delta_1,\Delta_2,B_1\otimes B_2}$ in $S'$ achieves the same effect as the substitution of $N_1$ and $N_2$ in $S'_1$. Hence the claim is proved.
We get the statement by applying the claim to $S$ and 
$\prfnet{A_1^{\set{\epsilon}},\ldots,A_n^{\set{\epsilon}}}{\set{A_1,\ldots,A_n}}$. \qed

\medskip
We can collect the results obtained so far.

\begin{theorem} \label{SEQ-DR-PARSING}
The following are equivalent for a paraproof structure $S$:
\begin{enumerate}
\item $S$ is a paraproof net, i.e., is sequentializable,
\item $S$ satisfies the DR-criterion,
\item $S$ satisfies the strong Parsing criterion,
\item $S$ satisfies the weak Parsing criterion.
\end{enumerate}
\end{theorem}

\Proof We have proved $(1)\Rightarrow (2)\Rightarrow (3)$ and $(4)\Rightarrow (1)$, and 
$(3)\Rightarrow (4)$ is obvious. \qed

\medskip
These equivalences are easy to upgrade to ``full MLL'', that is, to structures that also contain cuts.
We briefly explain how the definitions are extended. A paraproof structure can now be formalized as $\prfnetc{X'}{A_1^{U_1},\ldots,A_n^{U_n}}{X}$, where $X'$ is a (possibly empty) collection of disjoint subsets of $\set{A_1,\ldots,A_n}$, of the form $\set{B,B^\bot}$.  The conclusions of the paraproof structure are the 
formulas in $\set{A_1,\ldots,A_n}\setminus\Union X'$. The underlying graph is defined as above, 
with now in addition an edge between $B$ and $B^\bot$ for each class of the partial partition $X'$.
A paraproof is obtained 
as previously, with a new proof rule, the cut rule:
$$\seq{\vdash \Gamma_1,B\quad\vdash\Gamma_2,B^\bot}{\vdash\Gamma_1,\Gamma_2}$$

\noindent
A sequentializable paraproof structure is now one which is obtained through the following formal system, which adapts and extends (last rule) the one in Definition \ref{NET-criterion}.

$$\begin{array}{c}
\seq{}{\prfnetc{}{A_1^{\set{\epsilon}},\ldots,A_n^{\set{\epsilon}}}{\set{\set{A_1,\ldots,A_n}}}}\quad\quad
\seq{\prfnetc{X'}{A_1^{U_1},\ldots,A_n^{U_n},B^V,C^W}{X}}{\prfnetc{X'}{A_1^{U_1},\ldots,A_n^{U_n},(B\lpar C)^U}{X}}\\\\
\seq{\prfnetc{X'}{A_1^{U_1},\ldots,A_n^{U_n},B^V}{X}\;\;
\prfnetc{Y'}{{A'}_1^{U'_1},\ldots,{A'}_{n'}^{U'_{n'}},C^W}{Y}}{\prfnetc{X'\union Y'}{A_1^{U_1},\ldots,A_n^{U_n},{A'}_1^{U'_1},\ldots,{A'}_{n'}^{U'_{n'}},(B\otimes C)^U}{X\union Y}}\\\\
\seq{\prfnetc{X'}{A_1^{U_1},\ldots,A_n^{U_n},B^V}{X}\;\;
\prfnetc{Y'}{{A'}_1^{U'_1},\ldots,{A'}_{n'}^{U'_{n'}},(B^\bot)^W}{Y}}{\prfnetc{X'\union Y'\union\set{\set{B,B^\bot}}}{A_1^{U_1},\ldots,A_n^{U_n},{A'}_1^{U'_1},\ldots,{A'}_{n'}^{U'_{n'}},B^V,(B^\bot)^W}{X\union Y}}
\end{array}$$

\medskip\noindent
The parsing rewriting system is adapted and extended as follows:

$$\begin{array}{c}
\seq{A/u=B_1\lpar B_2}{\prfnetc{X'}{\Gamma,A^{U\union\set{u1,u2}}}{X,\set{\Delta,B_1,B_2}} 
\rightarrow_P
\prfnetc{X'}{\Gamma,A^{U\union\set{u}}}{X,\set{\Delta,(B_1\lpar B_2)}}}\\\\
\seq{A/u=B_1\otimes B_2}{\prfnetc{X'}{\Gamma,A^{U\union\set{u1,u2}}}{X,\set{\Delta_1,B_1},\set{\Delta_2,B_2}}   \rightarrow_P
\prfnetc{X'}{\Gamma,A^{U\union\set{u}}}{X,\set{\Delta_1,\Delta_2,B_1\otimes B_2}}}\\\\
\prfnetc{X',\set{B,B^\bot}}{\Gamma,B^\epsilon,(B^\bot)^\epsilon}{X,\set{\Delta_1,B},\set{\Delta_2,B^\bot}}
  \rightarrow_P
\prfnetc{X'}{\Gamma}{X,\set{\Delta_1,\Delta_2}}
\end{array}$$

\noindent
Of course, the formalization begins to become quite heavy. 
We  encourage the reader to draw the corresponding graph transformations.
Theorem  \ref{SEQ-DR-PARSING} holds for paraproof structures with cuts. The cut rule is
very similar to the $\otimes$ rule, and indeed the case of a cut in the proofs is treated exactly like that of the $\otimes$ rule. 

\medskip We have yet one more caracterization of (cut-free, this time) paraproof structures ahead, 
but let us pause and take profit from cuts: once they are in the picture, we want to explain how
to compute with them, that is, how to eliminate them. There is a single cut elimination rule, which transforms a cut between $B_1\otimes B_2$ and $B_1^\bot\lpar B_2^\bot$ into two  cuts
between $B_1$ and $B_1^\bot$, and between $B_2$ and $B_2^\bot$, and eliminates the
vertices  $B_1\otimes B_2$ and $B_1^\bot\lpar B_2^\bot$. Formally:

$$\begin{array}{l}
\prfnetc{X',\set{B_1\otimes B_2,B_1^\bot\lpar B_2^\bot}}
{\Gamma,(B_1\otimes B_2)^U,(B_1^\bot\lpar B_2^\bot)^V}
{X} \\
\quad\rightarrow\quad
\prfnetc{X',\set{B_1,B_1^\bot},\set{B_2,B_2^\bot}}
{\Gamma, B_1^{U_1},B_2^{U_2},(B_1^\bot)^{V_1},(B_2^\bot)^{V_2}}{X}
\end{array}$$

\noindent where $U_1=\setc{u}{1u\in U}$ and $U_2,V_1,V_2$ are defined similarly.
Let us stress the difference in nature between  cut-elimination ($\rightarrow$)  and parsing ($\rightarrow_P$): the former is of a dynamic nature (we compute something), while the latter is of a static nature (we check the correctness of something).

How are we sure that the resulting structure is a paraproof net, if we started from a paraproof net? This must be proved, and we do it next.

\begin{proposition} \label{cut-DR}
If  $S$ is a paraproof net and $S\rightarrow S'$, then $S'$ is also a paraproof net. 
\end{proposition}

\Proof In this proof, we use the second part of Proposition \ref{strong-DR}. 
Let us fix a switching for $S'$, and a switching of the $\lpar$ eliminated by the cut rule, say, $L$.
This induces a correction graph on $S$, which can be represented with $B_1^\bot\lpar B_2^\bot$ as root, with two sons $B_1\otimes B_2$ and $B_1^\bot$, where the former has in turn two sons $B_1$ and $B_2$. Let us call $T_1,T_2$ and $T'_1$ the trees rooted at $B_1,B_2$ and $B_1^\bot$, respectively.  The correction graph for $S'$ is obtained by placing 
$T_1$ as a new immediate subtree of $T'_1$ and $T_2$ as a new immediate subtree of the tree rooted at $B_2^\bot$. We use here the fact that we know that $B_2^\bot$ occurs in $T'_1$ (the important point is to make sure that it does not occur in $T_2$, as adding a cut  between 
$B_2$ and $B_2^\bot$ would result both in a cycle and in disconnectedness). This shows that $S'$ satisfies the DR-criterion, and hence is a paraproof net.
\qed

\medskip
We now describe yet another characterization  of paraproof nets,
elaborating on a suggestion of Girard in
\cite{GirMin1}. This last criterion is interactive in nature, and as such is an anticipation on the style of things that will be studied in the subsequent sections. The following definition is taken from
\cite{DanosTh,DR89} (where it is used to generalize the notion of multiplicative connective).

\begin{definition} Let $E$ be a finite set. Any two partitions $X$ and $Y$ of $E$ induce  a (bipartite) graph defined as follows: the vertices are the classes of $X$ and of $Y$,
the edges are the elements of $E$ and  for each $e\in E$, $e$ connects the class of $e$ in $X$ with the class of $e$ in $Y$. We say that $X$ and $Y$ are orthogonal if the induced graph is connected and acyclic.
\end{definition} 

Let $S$ be a paraproof structure with conclusions
$A_1^{U_1}, \ldots, A_n^{U_n}$, and consider a multiset  of paraproof nets 
$N_1=\prfnet{(A_1^\bot)^{U_1}}{X_1},\ldots,N_n=\prfnet{(A_1^\bot)^{U_n}}{X_n}$, which we
call collectively a counter-proof for $S$. By taking  $X_1\union\ldots\union X_n$,
we get a partition of the (duals of the) leaves of $S$, which we call the partition induced by the counter-proof.
We are now ready to formulate the Counterproof criterion, or CP-criterion for short.

\begin{definition}[CP]  Let $S$ be a paraproof structure. We say that $S$ satisfies the {\em CP-criterion} if its partition is orthogonal to all the partitions induced by all the counter-proofs of $S$.
\end{definition}

\begin{proposition} \label{DR-implies-CP}
If a paraproof structure $S$ satisfies the DR criterion, then it also satisfies the CP-criterion.
\end{proposition}

\Proof Let $N_1,\ldots,N_n$ be a counter-proof of $S$. We can form a paraproof net by placing 
cuts between $A_1$ and $A_1^\bot$, etc..., which also satisfies the DR criterion.
It is obvious to see that the cut-elimination ends exactly with the graph induced by the two partitions, and hence we conclude by Proposition \ref{cut-DR}. \qed

\begin{proposition} \label{CP-implies-DR}
If a paraproof structure satisfies the CP-criterion, then it satisfies the DR-criterion.
\end{proposition}

\Proof Let $S=\prfnet{A_1^{U_1},\ldots,A_n^{U_n}}{X}$ be a paraproof structure satisfying the CP-criterion and let us fix a switching for $S$. To this switching we associate a multiset $\Pi$ of extreme paraproofs of $\vdash A_1^\bot,\ldots,\vdash A_n^\bot$, defined as follows: the $\otimes$ rules of the counter-proof
have the form 

$$\seq{\vdash \Gamma,B_1\quad\vdash B_2}{\vdash \Gamma,B_1\otimes B_2}\quad \;\;\seql{or}{}{}\quad
\seq{\vdash B_1\quad\vdash \Gamma,B_2}{\vdash \Gamma,B_1\otimes B_2}$$

\noindent 
according to whether the corresponding $\lpar$ of $S$ has its switch set to $L$ or to $R$, respectively. By performing a postorder traversal of the counter-proof, we determine a sequence
$S_1,\ldots, S_p$  of paraproof structures and a sequence $\Pi_1,\ldots,\Pi_p$  of  multisets of extreme paraproofs, such that  the roots of $\Pi_i$ determine a partition of the conclusions of $S_i$, for all $i$. 
We start with $S_1=S$ and $\Pi_1=\Pi$. We construct $\Pi_{i+1}$ by removing the last rule used on one of the trees of  $\Pi_i$, and 
$S_{i+1}$ by removing from $S_i$ the formula (and its incident edges) dual to the formula decomposed by this rule (this is cut elimination!). We stop at stage $p$ when $\Pi_p$ consists of the generalized axioms of $\Pi$ only, and when $S_p$ consists of the partition of $S$ only. 
 Let $G_i$ be the graph obtained by taking the union of  $S_i$ restricted according to the (induced) switching  and of the set of the roots of $\Pi_i$, and by adding edges between 
a conclusion $A$ of $S_i$ and a root $\vdash\Gamma$ of $\Pi_i$ if and only if $A^\bot$ occurs in $\Gamma$.
We show by induction on $p-i$ that $G_i$ is acyclic and connected. Then applying this to $i=p-1$ we obtain the statement, since $G_1$ is the correction graph associated to our switching to which one has added (harmless) edges from the conclusions of $S$ to new vertices $\vdash A_1^\bot,\ldots,\vdash A_n^\bot$, respectively.
The base case follows by our assumption that $S$ satisfies the CP-criterion: indeed, the graph $G_p$ is obtained from the graph $G$  induced by $X$ and the partition induced by $\Pi$ by inserting a new node in the middle of each of its edges, and such a transformation yields a tree from a tree.

\smallskip
Suppose that one goes from $G_i$ to $G_{i+1}$ by removing
$$\seq{\vdash\Gamma,B_1,B_2}{\vdash \Gamma,B_1\lpar B_2}$$
Then, setting $N_1=\:\vdash\Gamma,B_1,B_2$ and $N'_1=\:\vdash\Gamma, B_1\lpar B_2$, $G_i$ is obtained from $G_{i+1}$ by renaming $N_1$ as $N'_1$, by removing the two edges between 
$B_1^\bot$ and $N_1$ and between $B_2^\bot$ and $N_1$, by adding a new vertex $B_1^\bot\otimes B_2^\bot$ and three new edges linking $B_1^\bot\otimes B_2^\bot$ with 
$B_1^\bot$, $B_2^\bot$, and $N'_1$. Considering $G_{i+1}$ as a tree with root $N_1$, we obtain
$G_i$ by cutting the subtrees rooted at $B_1^\bot$ and $B_2^\bot$ and by removing the corresponding edges out of $N_1$, by adding a new son to $N_1$ and by regrafting the subtrees as the two sons of
this new vertex. This clearly results in a new tree. 

\smallskip
Suppose now that  one goes from $G_i$ to $G_{i+1}$ by removing, say
$$\seq{\vdash\Gamma,B_1\quad\vdash B_2}{\vdash \Gamma,B_1\otimes B_2}$$
Then, setting $N_1=\:\vdash\Gamma,B_1$ and $N'_1=\:\vdash\Gamma,B_1\otimes B_2$, $G_i$ is obtained from $G_{i+1}$ by renaming $N_1$ as $N'_1$, by removing the vertex $\vdash B_2$ and the two edges between 
$B_1^\bot$ and $N_1$ and between $B_2^\bot$ and $\vdash B_2$, and by adding a new vertex $B_1^\bot\lpar B_2^\bot$ and two new edges linking $B_1^\bot\lpar B_2^\bot$ with 
$B_1^\bot$ and $N'_1$. Considering $G_{i+1}$ as a tree with root $\vdash B_2$, we obtain
$G_i$ by cutting the root $\vdash B_2$ (which had a unique son) and by inserting a new node ``in the middle'' of the edge between $B_1^\bot$ and $N_1$, which makes a new tree. \qed

\medskip
Hence, we have proved the following theorem.

\begin{theorem}
The following are equivalent for a paraproof structure $S$:
\begin{enumerate}
\item $S$ satisfies the CP-criterion,
\item $S$ satisfies the DR-criterion.
\end{enumerate}
\end{theorem}

\medskip
But there is more to say about this characterization. Let us make the simplifying assumption that we work with paraproof structures with a single conclusion $A$ (note that any paraproof structure can be brought to this form by inserting enough final $\lpar$ nodes). Let us say that a paraproof structure of conclusion $A$ is {\em orthogonal} to a paraproof structure of conclusion $A^\bot$ if their partitions are orthogonal. 
Given a set $H$ of paraproof structures of conclusion $B$, we write $H^\bot$ for the set of 
paraproof structures of conclusion $B^\bot$ which are orthogonal to all the elements of $H$.
We have (where ``of'' is shorthand for ``of conclusion''):

$$\begin{array}{lll}\set{\mbox{paraproof nets of }A^\bot}^\bot & = &
\set{\mbox{paraproof nets of }A} \\
& = &
\set{\mbox{extreme paraproof nets of }A^\bot}^\bot\;.
\end{array}$$

\noindent 
Indeed, Proposition \ref{DR-implies-CP} and the proof of Proposition \ref{CP-implies-DR} say that:

$$\begin{array}{l}
\set{\mbox{paraproof nets of }A}\inc \set{\mbox{paraproof nets of }A^\bot}^\bot \mbox{ and
}\\
\set{\mbox{extreme paraproof nets of }A^\bot}^\bot \inc
\set{\mbox{paraproof nets of }A}\;,
\end{array}$$

\noindent 
respectively, and the two equalities follow since
$$\set{\mbox{paraproof nets of }A^\bot}^\bot \inc \set{\mbox{extreme paraproof nets of }A^\bot}^\bot\;.$$
We also have:

$$\begin{array}{llll}
\set{\mbox{paraproof nets of }A} & = & 
\set{\mbox{paraproof nets of }A}^{\bot\bot}\\
& = & \set{\mbox{extreme paraproof nets of }A} ^{\bot\bot}\;.
\end{array}$$

\noindent Indeed, using the above equalities, we have:

$$\begin{array}{lll}
\set{\mbox{paraproof nets of }A}^{\bot\bot} & = & \set{\mbox{extreme paraproof nets of }A}^{\bot\bot}\\
& = &
\set{\mbox{paraproof nets of }A^\bot}^\bot\\
& = &\set{\mbox{paraproof nets of }A^{\bot\bot}}\\
& = &  \set{\mbox{paraproof nets of }A} \;,
\end{array}$$

\noindent from which the conclusion follows.  Anticipating on a terminology introduced in section \ref{behaviours}, we thus have that the set of paraproof nets of $A$ forms a {\em behaviour}, which is 
generated by the set of extreme paraproof nets. The paraproof nets of conclusion $A$ are those paraproof structures that ``stand the test" against all the counter-proofs $A^\bot$, which can be thought of opponents. We refer to Exercise \ref{AJ-criterion} for another criterion of a game-theoretic flavour.

\medskip 
Putting the two theorems together, we have obtained three equivalent characterizations of sequentializable paraproof structures: the DR-criterion, the Parsing criterion, and the CP-criterion. What about the characterization of sequentializable proof structures? All these equivalences cut down to proof structures, thanks to the following easy proposition.

\begin{proposition}
A sequentializable paraproof structure which is a proof structure is also a sequentializable proof structure.
\end{proposition}

\Proof Notice that any application of a generalized axiom in the construction of a paraproof net
remains in the partition until the end of the construction. Hence the only possible axioms must be of the form $\set{C,C^\bot}$. \qed

\medskip
We end the section by sketching how proof nets for MELL are defined, that is, how the proof nets can be extended to exponentials. Recall from section 3 that cut-elimination involves duplication of (sub)proofs.
Therefore, one must be able to know exactly which parts of the proof net have to be copied.
To this aim, Girard introduced {\em boxes}, which record some sequentialization information on the net.
Boxes are introduced each time a promotion occurs. The promotion amounts to add a new node to the graph, corresponding to the formula $!A$ introduced by the rule. But at the same time, the
part of the proof net that represents the subproof of conclusion $\vdash ?\Gamma, A$ is placed in a box.
The formulas of $?\Gamma$ are called the auxiliary ports of the box, while $!A$ is called the principal port. We also add:

\begin{itemize}
\item  {\em contraction nodes}, labelled with a formula $?A$, which have exactly two incoming edges also  labelled with $?A$,
\item {\em dereliction nodes}, also labelled with a formula $?A$, which have only one incoming edge labelled with $A$,
\item {\em weakening nodes}, labelled with $?A$, with no incoming edge.
\end{itemize}

\noindent
The corresponding cut-elimination rules are:

\begin{itemize}
\item
 the dereliction rule, which removes a cut between
a dereliction node labelled $?A^\bot$ and a principal port $!A$ and removes the dereliction node and the box (and its principal port) and places a new cut between the sons $A^\bot$ of $?A^\bot$ and
$A$ of $!A$;
\item the contraction rule, which removes a cut between a contraction node labelled $?A^\bot$ and a principal port $!A$ and removes the contraction node and duplicates the box and places two new cuts between the two copies of $!A$ and the two sons of $?A^\bot$ and adds contraction nodes connecting the corresponding auxiliary ports of the two copies;
\item the weakening rule, which removes a cut  between a weakening node labelled $?A^\bot$ and a principal port $!A$ and removes the weakening node and erases the box except for its auxiliary ports which are turned into weakening nodes.
\end{itemize}

All commutative conversions of sequent calculus (cf. part I, section 3) but one disappear in proof nets.
The only one which remains is:
\begin{itemize}
\item  a rule which concerns the cut between an auxiliary port $?A^\bot$ of a box and the principal port $!A$ of a second box, and which lets the second box enter the first one, removing the auxiliary port 
$?A^\bot$, which is now cut inside the box.
\end{itemize}

\smallskip
These rules, and most notably the contraction rule, are  global, as opposed to the elimination rules for MLL, which involve only the cut and the immediately neighbouring nodes and edges. By performing copying in local, elementary steps, one could hope to copy only what is necessary for the computation to proceed, and continue to share subcomputations that do not depend on the specific context of a copy.
Such a framework has been proposed by Gonthier and his coauthors  in  two insigthful papers \cite{GAL92a,GAL92b} bridging L\'evy's optimality theory \cite{Levy80}, Lamping's implementation of this theory \cite{Lamping90},  and proof nets. A book length account of this work can be found in \cite{AsGuer98}. It is related
to an important research program, called the {\em geometry of interaction} (GOI) \cite{Gir89a,Regnier92,DanosTh}. A detailed survey would need another paper. Here we shall only sketch what GOI is about.
The idea is to look at the paths (in the graph-theoretic sense) in a given proof net and to sort out those which are 
``meaningful " computationally. A  cut edge is such a path, but more generally, a path between 
two formulas which will be linked by a cut later (that is, after some computation) is meaningful. Such paths may be termed {\em virtual} (a terminology introduced by Danos and Regnier \cite{DR93}) since they say something about 
the reduction of the proof net without actually reducing it. Several independent characterizations of meaningful paths have been given, and turn out to be all equivalent \cite{ADLR94,AsGuer98}:

\begin{itemize} 
\item {\em Legal} paths \cite{AsLan93}, whose definition draws from a careful study of the form of L\'evy's labels, which provide descriptions of the history of reductions in the $\lambda$-calculus and were instrumental in his theory of optimality. 
\item {\em Regular} paths. One assigns weights (taken in some suitable algebras) to all the edges of the proof net, and the regular paths are those which have a non-null composed weight.
\item {\em Persistent} paths. These are the paths that are preserved by all computations, i.e. that no reduction sequence can break apart. Typically, in a proof net containing a $\otimes/\lpar$ cut
$$\seq{\seq{A\quad B}{A\otimes B}\quad\quad\seq{A^\bot\quad B^\bot}{A^\bot\lpar B^\bot}}{}$$
a path going down to $A$ and then up through $B^\bot$ gets disconnected when the cut has been replaced by the two cuts on $A$ and $B$, and hence is not persistent.
\end{itemize}

\smallskip
Any further
information on proof nets should be sought for in \cite{Gir87,GirLLSS}, and for additive proof nets (which are not yet fully understood) in, say,  \cite{Tor03,Hug03}. We just mention a beautiful complexity result
of Guerrini:  it can be decided in (quasi-)linear time whether an MLL proof structure is a proof net \cite{Gue99}.

\begin{exercise}[MIX rule]  \label{acyclicity-criterion}
 We define the {\em Acylicity criterion} by removing the connectedness requirement in the DR-criterion. Show that this criterion characterizes the paraproof structures that can be sequentialized with the help of the following additional rule:

$$\seq{\prfnet{A_1^{U_1},\ldots,A_n^{U_n}}{X}\quad\prfnet{{A'}_1^{U'_1},\ldots,{A'}_{n'}^{U'_{n'}}}{Y}}{\prfnet{A_1^{U_1},\ldots,A_n^{U_n},{A'}_1^{U'_1},\ldots,{A'}_{n'}^{U'_{n'}}}{X\union Y}}$$

\noindent
that corresponds to adding the following rule to the sequent calculus of MLL, known as the MIX rule:

$$\seq{\vdash \Gamma\quad\quad\vdash\Delta}{\vdash \Gamma,\Delta}$$

\noindent (Hint: adapt correspondingly the Parsing criteria.)
\end{exercise}

\begin{exercise}[AJ-criterion]  \label{AJ-criterion}
In this exercise, we provide a criterion for cut-free proofs which is equivalent to the Acyclicity criterion (cf. Exercise \ref{acyclicity-criterion}) and thus characterizes MLL+MIX cut-free proof nets, due to Abramsky and Jagadeesan \cite{AJ94,Baillot99}. 
Let $S=\prfnet{A_1^{U_1},\ldots,A_n^{U_n}}{X}$ be a proof structure, and let $$M=\setc{m^+}{m=(u,i), i\in\set{1,\ldots,n},u\in U_i}\union\setc{m^-}{m=(u,i),i\in\set{1,\ldots,n},u\in U_i}\;.$$
($M$ is a set of moves, one of each polarity for each conclusion of an axiom -- think of these formulas as the atoms out of which the conclusions of the proof structure are built).  Let 
$U=\setc{(v,i)}{v\mbox{ is a prefix of some }u\in U_i}$.
Given $(v,i)\in U$ and a sequence $s\in M^\star$, we define $s\rst{(v,i)}$ as follows:
$$\epsilon\rst{(v,i)}=\epsilon\quad\quad (s\,(u,j)^\lbd)\rst{(v,i)}=\left\{\begin{array}{ll}
(s\rst{(v,i)})\,(u,j)^\lbd & \mbox{if } i=j \mbox{ and } v\mbox{ is a prefix of }u\\
s\rst{(v,i)} & \mbox{otherwise}
\end{array}\right.$$
where $\lbd\in\set{+,-}$.
A finite sequence $s=m_1^-m_2^+m_3^-\ldots$ is called a {\em play} if it satisfies the following three conditions:
\begin{enumerate}
\item
for all $(v,i)\in U$, $s\rst{(v,i)}$ is alternating,
\item
for all $(v,i)\in U$,  if  $A_i/v=B_1\otimes B_2$, then only Opponent can switch
$(v,i)$-components  in $s\rst{(v,i)}$, 
\item
for all $(v,i)\in U$,  if  $A_i/v=B_1\lpar B_2$,
then only Player can switch
$(v,i)$-components  in $s\rst{(v,i)}$,
\end{enumerate}
where component switching is defined as follows: Opponent (resp. Player) switches $(v,i)$-components if $s\rst{(v,i)}$ contains two consecutive moves
$(u,i)^-(w,i)^+$ (resp. $(u,i)^+(w,i)^-$) where $u=v1u'$ and $w=v2w'$, or $u=v2v'$ and $w=v1w'$.

We need a last ingredient to formulate the criterion. To the partition $X$ we associate a function (actually, a transposition) $\phi$ which maps $(u,i)$ to $(w,j)$ whenever $\set{A_i/u,A_j/w}\in X$.  

\smallskip
We say that $S$ satisfies the {\em AJ-criterion} if 
whenever $m_1^-\,\phi(m_1)^+\,m_2^-\,\phi(m_2)^+\,\ldots \,m_n^-$ is a play,
then
$m_1^-\,\phi(m_1)^+\,m_2^-\,\phi(m_2)^+\,\dots\, m_n^-\,\phi(m_n)^+$ is also a play.
(It says that $\phi$ considered as a strategy is winning, i.e. can always reply, still abiding to the ``rules of the game".) 

\smallskip Show that an MLL+MIX sequentializable proof structure satisfies the AJ-criterion and  that a proof structure satisfying the AJ-criterion satisfies the Acylicity criterion. (Hints:  (1) Show using a minimality argument that if there exists a switching giving rise to a cycle, the cycle can be chosen such that no two tensors visited by the cycle are prefix one of another. (2) Fix a tensor on the cycle (there must be one, why?), walk on the cycle going up from there, and let  $(u,i),\phi(u,i),\ldots,(w,j),\phi(w,j)$ be the sequence of axiom conclusions visited along the way.
Consider $s=(u,i)^-\,\phi(u,i)^+\,\ldots\,(w,j)^-\,\phi(w,j)^+$. Show that if the AJ-criterion is satisfied, then all the strict prefixes of $s$ are plays.)
Conclude that the AJ-criterion characterizes MLL+MIX proof nets.
\end{exercise}

\begin{exercise} \label{type-iso}
Reformulate the example of cut-elimination given in part I, section 3:
$$\seq{\stackrel{\vdots}{\vdash ?A^\bot\lpar?B^\bot,!(A\& B)}\quad\quad\quad\stackrel{\vdots}{\vdash !A\otimes !B,?(A^\bot\oplus B^\bot)}}{\vdash !(A\with B),?(A^\bot\oplus B^\bot)}$$
using proof nets instead of sequents. Which reductions rule(s) should be added in order to 
reduce this proof to (an $\eta$-expanded form of the) identity (cf. part I, Remark 3.1)? Same question for the elimination of the cut on $!(A\with B)$. We refer to \cite{Laurent04} for a complete study of type isomorphisms in (polarized) linear logic.
\end{exercise}

\section{Towards ludics} \label{ludics}

The rest of the paper is devoted to ludics, a new research program started by Girard in \cite{Gir01}.
Ludics arises from forgetting the logical contents of formulas, just keeping their {\em locative} structure, i.e., their shape as a tree, or as a storage structure.  For example, if $(A\lpar B)\oplus A$ becomes an address $\xi$, then the (instances of) the subformulas $A\lpar B$ (and its subformulas $A$ and $B$) and $A$ become $\xi 1$ (and $\xi11$ and $\xi 12$) and $\xi 2$, respectively. 
The relation between formulas and addresses is quite close to the relation between typed and untyped
$\lambda$-calculus, and this relation will be made clear in the following section (as a modest contribution of this paper). 

Thus, another aspect of resource consciousness comes in: not only the control on the use of resources, which is the main theme of linear logic, but also the control on their storage in a shared memory.  An important consequence  of this is that ludics  gives a logical status to subtyping, a feature
related to object-oriented programming languages. It has been long observed that the semantics of subtyping, record types, intersection types, is not categorical, i.e. cannot be modelled naturally in the framework of category theory, unlike the simply typed $\lambda$-calculus, whose match with cartesian closed categories has been one of the cornerstones of denotational semantics over the last twenty years. Ludics points to the weak point of categories: everything there is up to isomorphism, up to renaming. The framework of ludics forces us to explicitly recognize
that a product of two structures calls for the copying of their shape in two disjoint parts of the memory. When this is not done, we simply get intersection types, so the product, or additive conjunction, appears as a special case of a more general connective, called the intersection. 

As a matter of fact, prior to the times of categorical semantics, denotational semantics took models of the untyped $\lambda$-calculus as its object of study. The whole subject started with Scott's solution to the domain equation $D=D\rightarrow D$, which allows us to give a meaning to self-application. The early days of semantics were concerned with questions such as the completeness of type assignment to untyped terms (as happens in programming languages like ML which have static type inference) \cite{Hindley83,BCD83}. Types were interpreted as suitable subsets of the untyped model, the underlying intuition being that of types as properties: a type amounts to the collection of all (untyped) terms to which it can be assigned.  In this framework, subtyping is interpreted by inclusion, and intersection types by ... intersection \cite{Mitchell88, BruceLongo90}.
Ludics invites us to revisit  these ideas with new glasses, embodying locations, and interactivity: in ludics, the intuition is that of types as behaviours, which adds an interactive dimension to the old paradigm.


\medskip
Another key ingredient of ludics is the {\em daimon}, which the author of this paper  recognized as an  {\em error}  element.  We mean here a recoverable error in the sense of 
Cardelli-Wegner \cite{CW85}. Such an error is a rather radical form of output, that stops execution and gets propagated to the top level. If the language is endowed with error-handling facilities, then
programs can call other programs called handlers upon receiving such error messages, whence the name "recoverable".
The importance of errors in denotational semantics was first noted by Cartwright and Felleisen in (1991)
(see  \cite{CCF92}).
 If $A$ is an algorithm of two arguments such that  $A(err,\bot)=err$, this test, or {\em interaction} between $A$ and $(err,\bot)$ tells us that $A$ wants to know its first argument before anything else.
This information reveals us a part of the computation strategy of $A$.
Note that $A(err,\bot)=\bot$ holds for an algorithm $A$ that will examine its second argument first.
It may then seem that $err$ and $\bot$ play symmetric r\^oles. This is not quite true, because $\bot$ means ``undefined'', with the meaning of ``waiting for some information'', that might never come.
So $err$ is definitely terminating, while we don't know for $\bot$: $\bot$ is a sort of error whose meaning is overloaded with that of non-termination. 

In ludics, Girard introduces (independently) the {\em daimon}, denoted by $\demon$, with the following motivation. Ludics is an interactive account of logic. So the meaning of a (proof of a) formula $A$ lies in its behavior against observers, which are the proofs of
other formulas, among which the most essential one is its (linear) negation $A^\bot$ (not $A$, the negation of linear logic \cite{Gir87,GirLLSS}). This is just like the contextual observational semantics of a program, where the meaning of a piece of program $M$ is given by all the observations of the results of the evaluation of $C[M]$, where $C$ ranges over full (closed, basic type)  program contexts.
Now, there is a problem: if we have a proof of $A$, we hardly have a proof of $A^\bot$. There are not enough ``proofs'': let us admit more! This is the genesis of the daimon. Proofs are extended to paraproofs (cf. section \ref{proof-nets}):  in a paraproof, one can place daimons to signal that one gives up, i.e., that one assumes some formula instead of proving it. 
Hence there is always a paraproof  of $A^\bot$: the daimon itself, which stands for ``assume $A^\bot$'' without even attempting to start to write a proof of it. The interaction between any proof of $A$ and this ``proof'' reduced to the daimon terminates immediately. There are now enough inhabitants to 
define interaction properly!

We borrow the following comparison from Girard. In linear algebra, one expresses that two vectors $x$ and $y$ are orthogonal by writing that  their scalar product $\reseau{x}{y}$  is $0$: we are happy that there is ``enough space'' to allow a vector space and its orthogonal to have a non-empty intersection.
Like 0, the daimon inhabits all (interpretations of) formulas, and plays the role of an absorbing element
(as we shall see, it is orthogonal to all its counter-paraproofs).  It even inhabits -- and is the only (defined) inhabitant of --  the empty sequent, which in this comparison could be associated with the base field. But the comparison should not be taken too seriously.

In summary, errors or daimons help us to terminate computations 
and to explore the behavior of programs or proofs interactively.
Moreover, computation is  \guil{{\em streamlike}}, or {\em demand-driven}. The observer detains the prompt for further explorations.
If he wants to know more, he has to further defer giving up, and hence to display more of his own behavior. This is related to lazy style in programming, where one can program, say, the infinite list of prime numbers in such a way that each new call of the program will disclose the next prime number.  Coroutines come to mind here too, see \cite{CuSym01} for a discussion. 

\medskip A third ingredient of ludics is {\em focalization}, which we explain next.
In section 2 of part I, we observed that the connective $\otimes$ distributes over $\oplus$ and that both
connectives are irreversible, and that on the other hand $\lpar$ distributes over $\with$ and that both
connectives are reversible.  We introduced there the terminology of positive connectives  ($\otimes$, $\oplus$) and negative connectives ($\lpar$, $\with$). We extend the terminology to formulas as follows: a formula is positive (resp. negative) if its topmost connective is positive (resp. negative).

Andreoli shed further light on this division through his work on focalization  \cite{Andreoli92}. His motivation was to reduce the search space for proofs in linear logic.
Given, say, a positive formula, one groups its positive connectives from the root, and then  its negative connectives, etc.... Each grouping is considered as a single synthetic connective.  Let us illustrate this  with an example. Consider
$((N_1\otimes N_2)\with Q)\lpar R$, with $Q,R$, and $P=N_1\otimes N_2$ positive and $N_1$, $N_2$ negative. The signs, or {\em polarities} of these formulas show  evidence of the fact  that  maximal groupings of connectives of the same polarity have been made. We have two synthetic connectives, a negative and ternary one that associates $(P\with Q)\lpar R$ with  $P,Q,R$, and a positive one which is just the connective $\otimes$.  The rule for the negative synthetic  connective is:
$$\seq{\vdash P,R,\Lambda\quad\quad\vdash Q,R,\Lambda}{\vdash (P\with Q)\lpar R,\Lambda}$$
Thus, a  focused proof of $((N_1\otimes N_2)\with Q)\lpar R$ ends as follows:

$$\seq{\seq{\vdash N_1,R,\Lambda_1\quad \vdash N_2,\Lambda_2}{\vdash N_1\otimes N_2,R,\Lambda}\quad\quad\vdash Q,R,\Lambda}{\vdash ((N_1\otimes N_2)\with Q)\lpar R,\Lambda}$$

\noindent
Notice the alternation of the active formulas in the proof: negative ($(P\with Q)\lpar R$), then positive ($N_1\otimes N_2$), etc...   (We recall that  at each step the active formula is the formula whose topmost connective has just been introduced.)
The same proof can be alternatively presented as follows:

$$\seq{\seq{N_1^\bot \vdash R,\Lambda_1\quad N_2^\bot \vdash \Lambda_2}{\vdash N_1\otimes N_2,R,\Lambda}\quad\quad\vdash Q,R,\Lambda}{((N_1\otimes N_2)^\bot\oplus Q^\bot)\otimes R^\bot\vdash \Lambda}$$

\noindent 
The advantage of this formulation is that it displays only positive connectives. Notice also that
there is at most one formula on the left of $\vdash$. This   property is  an invariant of focalization,
and is a consequence of the following observation.

\begin{remark} \label{un-seul-a-gauche}
In a bottom-up reading of the rules of MALL, only  the $\lpar$ rule augments the number of formulas in the sequent. In  the other three rules for $\with$, $\otimes$, and $\oplus$, the active formula is replaced by a single subformula, and the context stays the same or gets shrinked. Thus, only a negative formula can give rise to more than one formula in the same sequent when it is active. 
\end{remark}

\noindent
Adapting this remark to synthetic connectives, we see that only a negative synthetic connective can give rise to more than one formula when it is active, and these formulas are positive by the maximal grouping.  

\smallskip
The {\em focusing discipline} is defined as follows:
\begin{enumerate}
\item Once the proof-search starts the decomposition of a formula, it keeps decomposing its subformulas until a connective of the opposite polarity is met, that is, the proof uses the rules for synthetic connectives;
\item A negative formula if any  is given priority for decomposition.
\end{enumerate}

The focusing discipline preserves the invariant that a (monolateral) sequent contains at most one negative formula. Indeed, if the active formula is positive, then all the other formulas in the sequent are also positive (as if there existed a negative formula it would have to be the active one), and then all the premises in the rule have exactly one negative formula, each of which arises from the decomposition of the active formula (like $N_1$ or $N_2$ above). If the active formula is negative, then all the other formulas of the sequent are positive, and each premise of the rule is a sequent consisting of positive formulas only (cf. $P, Q, R$ above).
Initially, one wants to prove a sequent consisting of a single formula, and such a sequent satisfies the invariant.

From now on, we consider sequents consisting of positive formulas only and with at most one formula on the left of $\vdash$. We only have positive synthetic connectives, but we now have a left rule and a set of right rules for each of them: the right rules are irreversible, while the left rule is reversible. Note that the left rule is just a reformulation of the right rule of the corresponding negative synthetic connective. Here are the rules for the ternary  connective $(P^\bot\oplus Q^\bot)\otimes R^\bot$:

$$\begin{array}{lll}
\seql{$\set{\set{P,R},\set{Q,R}}$}{\vdash P,R,\Lambda\quad\quad\vdash Q,R,\Lambda}{(P^\bot\oplus Q^\bot)\otimes R^\bot\vdash \Lambda} &&
\begin{array}{l}\seql{$\set{P,R}$}{P\vdash\Gamma\quad\quad R\vdash\Delta}{\vdash (P^\bot\oplus Q^\bot)\otimes R^\bot,\Gamma,\Delta}\\\\
\seql{$\set{Q,R}$}{Q\vdash\Gamma\quad\quad R\vdash\Delta}{\vdash (P^\bot\oplus Q^\bot)\otimes R^\bot,\Gamma,\Delta}
\end{array}\end{array}$$

\noindent
Here is how the top right rule has been synthesized:

$$\seq{\seq{P\vdash \Gamma}{\vdash P^\bot\oplus Q^\bot,\Gamma}\quad\quad R\vdash \Delta}{\vdash (P^\bot\oplus Q^\bot)\otimes R^\bot,\Gamma,\Delta}$$

\noindent
The synthesis of the other rules is similar. Note that the isomorphic connective $(P^\bot\otimes Q^\bot)\oplus (P^\bot\otimes R^\bot)$, considered as a ternary connective, gives rise to the same rules. For example the top right rule is now synthesized as follows:

$$\seq{\seq{P\vdash \Gamma\quad R\vdash\Delta}{\vdash P^\bot\otimes R^\bot,\Gamma}}{\vdash ((P^\bot\otimes Q^\bot)\oplus (P^\bot\otimes R^\bot),\Gamma,\Delta}$$

\smallskip
More generally, it is easily seen that any positive synthetic connective, viewed as a linear term
over the connectives $\otimes$ and $\oplus$, can be written as a $\oplus$ of $\otimes$, modulo distributivity and associativity:
$$P(N_1,\ldots,N_k)=\cdots\; \oplus (N_{\phi(m,1)}\otimes \cdots\otimes N_{\phi(m,j_m)})\oplus\;\cdots$$ 

\noindent where $m$ ranges over $\set{1,\ldots n}$ for some $n$,  the range of $\phi$ is
$\set{1,\ldots,k}$, for each $m$ the map $\phi(m,\_)=\lbd j.\phi(m,j)$ is injective, and the map $\lbd m.\phi(m,\_)$ is injective (for example, for  $(N_1\oplus N_2)\otimes N_3$, we have $n=2$, $j_1=j_2=2$, $\phi(1,1)=1$, $\phi(1,2)=3$, $\phi(2,1)=2$, and $\phi(2,2)=3$). Note that the connective is equivalently described by a (finite) set of finite subsets of $\set{1,\ldots,k}$.  
The rules for $P(\cdots)$ are as follows (one rule for each $m\in\set{1,\ldots n}$):

$$\seq{N_{\phi(m,1)}\vdash\Lambda_1\quad\cdots\quad N_{\phi(m,j_m)}\vdash\Lambda_{j_m}}{\vdash P(N_1,\ldots,N_k),\Lambda_1,\ldots\Lambda_{j_m}}
$$

\noindent
There is one rule (scheme) corresponding to each component of the $\oplus$ that selects this component and splits its $\otimes$. Dually, a negative synthetic connective can be written as a $\with$ of $\lpar$:
$$N(P_1,\ldots,P_k)=\cdots\; \with\; (P_{\phi(m,1)}\lpar \cdots\lpar P_{\phi(m,j_m)})\;\with\;\cdots$$ 

\noindent
Here is the rule for $N(\cdots)$.

$$\seq{\vdash P_{\phi(1,1)},\ldots,P_{\phi(1,j_1)} ,\Lambda\quad\cdots\quad\vdash P_{\phi(n,1)},\ldots,P_{\phi(n,j_n)} ,\Lambda}{N(P_1,\ldots,P_k)^\bot\vdash\Lambda}$$

\noindent
There is one premise corresponding to each component of the $\with$, and in each premise the $\lpar$'s have been dissolved into a sequence of positive formulas.

\begin{remark} \label{neg-pas-choix}
Note that in the  negative rule we have no choice for the active formula: it is the unique formula on the left. Thus, the negative rule is not only reversible at the level of the formula $N(\cdots)$, but also at the level of the sequent $N(\cdots)^\bot\vdash\Lambda$. In contrast, in a  positive rule, one has to choose not only the $m$ and the disjoint $\Lambda_1,\ldots,\Lambda_{j_m}$, as noted above, but also the formula $P(\cdots)$ in the sequent $\vdash P(\cdots),\Lambda$.
\end{remark}

Below, we display some examples of non-focusing proofs:

$$\seq{\seq{\seq{\vdots}{\vdash P\with Q,N_1, \Lambda_1}
\quad \seq{\vdots}{\vdash R, N_2,\Lambda_2}}
{\vdash P\with Q, R,N_1\otimes N_2, \Lambda_1,\Lambda_2}}
{\vdash (P\with Q)\lpar R,N_1\otimes N_2, \Lambda_1,\Lambda_2}$$

$$\seq{\seq{\vdots}{\vdash (P\with Q)\lpar R,N_1, \Lambda_1}
\quad \seq{\vdots}{ \vdash  N_2,\Lambda_2}}
{\vdash (P\with Q)\lpar R,N_1\otimes N_2, \Lambda_1,\Lambda_2}$$

\noindent
In the first proof, we did not respect the maximal grouping of negative connectives, and we abandoned our focus on $(P\with Q)\lpar R$ to begin to work on another formula of the sequent. In the second proof, we did not respect the priority for the negative formula of the sequent.
These examples show that the 
change of granularity induced by focusing is not innocuous, because the focusing discipline forbids some proofs. Of course, this was the whole point for Andreoli and Pareschi, who wanted to reduce the search space for proofs.  But is the focusing discipline complete in terms of provability? The answer is yes \cite{Andreoli92}, i.e., no provable sequents are lost. The result actually holds for the whole of linear logic, with $!$ (resp. $?$) acting on a negative (resp. positive) formula to yield a positive (resp. negative) formula (see Remark \ref{sharp-flat}).

\medskip
Before we state and sketch the proof of the focalization theorem, we introduce a focusing sequent calculus \cite{Andreoli92,LS00} which accepts only the focusing proofs.
First, we note that a better setting for polarized formulas consists in insisting that a positive connective should connect positive formulas and a negative connective should connect negative formulas. This is possible if we make changes of polarities explicit with the help of change-of-polarity operators (cf., e.g., \cite{Laurent02}).  For example, $((N_1\otimes N_2)\with Q)\lpar R$ should be written as $(\uparrow\! ((\downarrow\! N_1))\otimes \!(\downarrow\! N_2))\with \!(\uparrow\! Q))\lpar \!(\uparrow\! R)$. For MALL, we get the following syntax for
positive and negative formulas (assuming by convention that the atoms $X$ are all positive, which is no loss of generality since we get ``as many" negative atoms $X^\bot$).

$$\begin{array}{l}
P::= X \Alt P\otimes P \Alt P\oplus P \Alt 1 \Alt 0 \Alt \makep{N}\\
N::= X^\bot \Alt N\lpar N\Alt N\with N\Alt \bot\Alt \top\Alt \maken{P}
\end{array}$$

\noindent
The operators $\makep{}$ and $\maken{}$ are called the {\em shift} operations.
There are two kinds of sequents: $\stoup{\Gamma}{}$ and $\stoup{\Gamma}{P}$
where in the latter the only negative formulas allowed in $\Delta$ are atoms. The zone in the sequent on the right of ``;" is called the {\em stoup}, a terminology which goes back to an earlier paper of Girard on classical logic
\cite{GirLC}. The stoup is thus either empty or contains exactly one formula.
The rules of this sequent calculus are as follows (we omit the units):

$$\begin{array}{llclc}
&& \seq{}{\stoup{P^\bot}{P}} && \seql{\mbox{\bf Focalization}}{\stoup{\Gamma}{P}}{\stoup{\Gamma,P}{}}\\\\
\mbox{\bf SHIFT}\\
&& \seq{\stoup{\Gamma,P}{}}{\stoup{\Gamma,\maken{P}}{}} &&
\seq{\stoup{\Delta,N}{}}{\stoup{\Delta}{\makep{N}}}\\\\
\mbox{\bf POSITIVE}\\
 &&  \seq{\stoup{\Gamma}{P_1}}{\stoup{\Gamma}{P_1\oplus P_2}} \quad \seq{\stoup{\Gamma}{P_2}}{\stoup{\Gamma}{P_1\oplus P_2}} &&
\seq{\stoup{\Gamma_1}{P_1}\quad\stoup{\Gamma_2}{P_2}}{\stoup{\Gamma_1,\Gamma_2}{P_1\otimes P_2}}
 \\\\
\mbox{\bf NEGATIVE}\\
&& \seq{\stoup{ N_1,N_2, \Gamma}{}}{\stoup{ N_1\lpar N_2, \Gamma}{}} &&
\seq{\stoup{ N_1,\Gamma}{}\quad\stoup{N_2,\Gamma}{}}{\stoup{ N_1\& N_2,\Gamma}{}}
\end{array}$$
 
\smallskip \noindent
 In the focalization rule we require that the negative formulas of $\Gamma$ (if any) are atomic.
 It should be clear that focusing proofs are in exact  correspondence with the proofs of linear logic that respect the focusing discipline. In one direction, one inserts the shift operations at the places where
 polarity changes, like we have shown on a sample formula, and in the other direction one just forgets these operations.  The synthetic connective view further abstracts from the order in which the negative subformulas of a negative connective are decomposed.
 
 \begin{remark} \label{weak-focalization} In view of the above sequent calculus, a weaker focusing discipline appears more natural, namely that obtained by removing the constraint that in a sequent
 with a non-empty stoup the context consists of positive formulas and negative atoms only. What then remains of the focusing discipline  is to maintain the focus on the decomposition of {\em positive} formulas, which is enforced under the control of the stoup.
 \end{remark}
 
\begin{theorem}[Focalization] The focusing discipline is complete, i.e.,  if  $\:\vdash A$ is provable in linear logic, then it is provable by a cut-free proof respecting the focusing discipline.
\end{theorem}

\Proofhint The theorem says among other things that nothing is lost by giving priority to the negative connectives. In the two examples of non-focusing proofs, the decomposition of  the $\otimes$ can be easily permuted with the decomposition of the $\with$, or of the $\lpar$. Conversely, the decomposition of $\lpar$ has given more possibilities of proofs for the decomposition of $\otimes$, allowing to send $P\with Q$ on the left and $R$ on the right. But, more importantly, the theorem also says that nothing is lost by focusing on a positive formula and its topmost positive subformulas. More precisely, not any positive formula of the sequent to prove will do, but at least one of them can be focused on, as the following example shows:

$$\seq{\seq{\seq{\seq{\seq{}{\vdash N_2,N_2^\bot}\quad\seq{}{\vdash N_3,N_3^\bot}}{\vdash N_2\otimes N_3, N_2^\bot, N_3^\bot}}{\vdash N_2\otimes N_3, N_2^\bot\lpar N_3^\bot}\quad\quad\seq{}{\vdash N_4,N_4^\bot}}{\vdash N_2\otimes N_3, N_4\otimes(N_2^\bot\lpar N_3^\bot), N_4^\bot}}{\vdash N_1\oplus(N_2\otimes N_3), N_4\otimes(N_2^\bot\lpar N_3^\bot), N_4^\bot}$$

\smallskip\noindent
This proof is not focused:  the decomposition of the first positive formula $N_1\oplus(N_2\otimes N_3)$ of the conclusion  sequent is blocked, because it needs to access negative subformulas of the second positive formula. But focusing on the second formula works:

$$\seq{\seq{\seq{\seq{\seq{}{\vdash N_2,N_2^\bot}\quad\seq{}{\vdash N_3,N_3^\bot}}{\vdash N_2\otimes N_3, N_2^\bot,N_3^\bot}}{\vdash N_1\oplus(N_2\otimes N_3), N_2^\bot,N_3^\bot}}{\vdash N_1\oplus(N_2\otimes N_3), N_2^\bot\lpar N_3^\bot}\quad\quad\seq{}{\vdash N_4,N_4^\bot}}{\vdash N_1\oplus(N_2\otimes N_3), N_4\otimes(N_2^\bot\lpar N_3^\bot), N_4^\bot}$$

\smallskip
A simple proof of the focalization theorem can be obtained
by putting these observations together (Saurin \cite{Saurin04}). More precisely, the following 
properties of sequents of linear logic are easily proved:

\begin{enumerate}
\item[(1)] Any provable sequent $\vdash\Gamma,N$ has a proof in which $N$ is active at the last
step: this is an easy consequence of the fact that a configuration ``negative rule followed by positive rule" can be permuted to a configuration ``positive rule followed by negative rule" (and not conversely).

\item[(2)] Any provable sequent $\vdash\Gamma$ consisting of positive formulas only is such that there exists at least
one formula $P$ of $\Gamma$ and a proof of $\vdash\Gamma$ which starts (from the root) with a complete decomposition of  (the synthetic connective underlying) $P$. Let us sketch the proof. If the last step of the proof is a $\oplus$ introduction, say, $\Gamma=\Delta,A_1\oplus A_2$  and $\vdash\Gamma$ follows from $\vdash\Delta,A_1$,
induction applied to the latter sequent yields either $A_1$, in which case $A_1\oplus A_2$ will do, or
a formula $P$ of $\Delta$, in which case we take that formula, and we modify the proof of $\vdash\Delta,A_1$ obtained by induction, as follows: we spot the place where $A_1$ appears when the decomposition of $P$ is completed, and insert the $\oplus$ introduction there, replacing systematically $A_1$ by
$A_1\oplus A_2$ from the insertion point down to the root, yielding a proof of $\vdash\Delta,A_1\oplus A_2$
that satisfies the statement.  The tensor case is similar. 
\end{enumerate}

Then a focalized proof can be obtained out of any MALL proof by repeatedly applying (1) to a cut-free proof so as to
turn the sequent into a sequent of positive formulas only, and then (2), etc... The size of the sequents decreases at each step, so the procedure creates no infinite branch. The same property is likely to work in presence of exponentials, using the same kind of ordinal as in a cut-elimination proof.

\medskip
Steps (1) and (2) may seem ad hoc manipulations, but in fact, they can be taken care of
by cut-elimination. 
For example, setting $R'=P\with Q$,  we can reshape the proof 

$$\seq{\seq{\vdots}{\vdash R'\lpar R,N_1, \Lambda_1}
\quad \seq{\vdots}{\vdash  N_2,\Lambda_2}}
{\vdash R'\lpar R,N_1\otimes N_2, \Lambda_1,\Lambda_2}$$

\noindent
so as to get a proof ending with the decomposition of the negative connective $\lpar$, as follows:

$$\seq{\seq{ \seq{\seq{\vdots}{\vdash R'\lpar R,N_1, \Lambda_1}\quad\seq{\seq{}{\vdash  R'^\bot,R'}\quad\seq{}{\vdash R^\bot, R}}{\vdash (R'\lpar R)^\bot,R,R'}}{\vdash R', R,N_1, \Lambda_1}
\quad \seq{\vdots}{\vdash  N_2,\Lambda_2}}{\vdash R', R,N_1\otimes N_2, \Lambda_1,\Lambda_2}}{\vdash R'\lpar R,N_1\otimes N_2, \Lambda_1,\Lambda_2}$$

\noindent This kind of transformation can be performed systematically, so
 as to yield a proof which can be formalized in the focused sequent calculus {\em with cuts}.
 The resulting proof can then be run through the cut-elimination in the focused caclulus,
 yielding a focused and cut-free proof. This approach is closer to Andreoli's original proof  (see \cite{LauFoc} for details). \qed


\begin{remark} \label{sharp-flat}
As a hint of how the focalization theorem extends to exponentials, let us mention that
Girard has proposed to decompose $!N$ and $?P$ as $\makep{\sharp N}$ and $\maken{\flat P}$ , where $\sharp$ is a negative connective, taking a negative formula $N$ to a negative formula $\sharp N$ and where $\flat$ is dually a positive connective. The change-of-polarity aspect of $!$ and $?$ is
taken care of by the already introduced just-change-of-polarity operators $\makep{}$ and $\maken{}$.
The rules for these connectives in the focusing system are as follows:

$$\seq{\stoup{?\Gamma,N}{}}{\stoup{?\Gamma,\sharp N}{}}\quad\quad
\seq{\stoup{\Gamma}{P}}{\stoup{\Gamma}{\flat P}}\quad\quad
\seq{\stoup{\Gamma}{}}{\stoup{\Gamma,\flat P}{}}\quad\quad
\seq{\stoup{\Gamma,\flat P,\flat P}{}}{\stoup{\Gamma, \flat P}{}}$$
\end{remark}

\begin{exercise}
Show that the focusing proof system (in its weaker form) enjoys the cut-elimination property.
\end{exercise}

\begin{exercise}
Show that $\sharp$ is reversible, and discuss the irreversibility of $\flat$.
\end{exercise}

\section{Designs} \label{designs}

We now arrive at the basic objects of ludics: the {\em designs}, which are  
 ``untyped paraproofs" (cf. section \ref{proof-nets}). Designs are (incomplete) proofs
from which the logical content has been almost erased. Formulas are replaced by their (absolute) addresses, which are sequences of relative addresses recorded as numbers. Thus addresses are words of natural numbers.
We let $\zeta, \xi,\cdots \in\omega^\star$ range over addresses. We use $\xi\star J$ to denote $\setc{\xi j}{j\in J}$, and we sometimes write $\xi\star i$ for $\xi i$. The empty address is written  $\epsilon$. 
Girard uses the terminology {\em locus} and {\em bias} for absolute address and relative address, respectively. Let us apply this forgetful transformation to our example $(P^\bot\oplus Q^\bot)\otimes R^\bot$  of the previous section, assigning relative addresses $1,2,3$ to $P^\bot,Q^\bot,R^\bot$, respectively.

$$\begin{array}{lll}
\seql{$(-,\xi,\set{\set{1,3},\set{2,3}})$}{\vdash \xi 1,\xi 3,\Lambda\quad\quad\vdash \xi 2,\xi 3,\Lambda}{\xi\vdash \Lambda} &&
\begin{array}{l}\seql{$(+,\xi,\set{1,3})$}{\xi 1\vdash\Gamma\quad\quad \xi 3\vdash\Delta}{\vdash \xi,\Gamma,\Delta}\\\\
\seql{$(+,\xi,\set{2,3})$}{\xi 2\vdash\Gamma\quad\quad \xi 3\vdash\Delta}{\vdash\xi, \Gamma,\Delta}
\end{array}\end{array}$$

\noindent
Notice the sets $\set{1,3}$ and $\set{2,3}$ appearing in the names of the rules. They 
can be thought of as a reservation of some (relative) addresses, to be used to
 store immediate subformulas of the current formula. Girard calls these sets {\em ramifications}.
 

\smallskip
We then consider a notion of \guil{abstract sequent}: a negative (resp. positive)  {\em base} (also called {\em pitchfork} by Girard) has the form
$\xi\vdash\Lambda$ (resp. $\vdash \Lambda$), where $\Lambda$ is a finite set of addresses, i.e., a base is a sequent of addresses. We always assume a {\em well-formedness} conditions on bases: that all addresses in a base are pairwise disjoint
(i.e. are not prefix one of the other). We spell out this condition in the key definitions.

\medskip
The {\em designs} are trees  built via the following rules ($\omega$ denotes the set of natural numbers, and ${\cal P}_f(\omega)$ denotes the set of finite subsets of $\omega$):

\medskip
\noindent {\bf Daimon}: ($\vdash\Lambda$ well formed)

$$\seql{$\demon$}{}{\vdash\Lambda}$$

\noindent
{\bf Positive rule}  ($I\inc \omega$ finite, one premise for each $i\in I$,  all $\Lambda_i$'s pairwise disjoint and included in $\Lambda$, $\vdash\xi,\Lambda$ well formed):

$$\seql{$(+,\xi,I)$}{\cdots\quad\xi i\vdash\Lambda_i\quad\cdots}{\vdash\xi,\Lambda}$$

\noindent
{\bf Negative rule} (${\cal N}\inc{\cal P}_f(\omega)$ possibly infinite, one premise for each $J\in {\cal N}$, all $\Lambda_J$'s  included in $\Lambda$, $\xi\vdash\Lambda$ well formed):

$$\seql{$(-,\xi,{\cal N})$}{\cdots\quad\vdash \xi\star J,\Lambda_J\cdots}{\xi\vdash\Lambda}$$

\noindent  The first rule is the type-free version of the generalized axioms of  Definition 
\ref{paraproof-structure},
while the two other rules are the type-free versions of the rules for the synthetic connectives given in section \ref{ludics}.
The root of the tree is called the {\em base} of the design.
A design is called negative or positive according to whether its base is positive or negative. 

\medskip
Here is how a  generic negative design ends.

$$\seql{$(-,\xi_1,{\cal N}_1)$}{\cdots\;\; \seql{$(+,\xi_2,I_1)$}{\cdots\;\;\seql{$(-,\xi_2i_1,{\cal N}_2)$}{\cdots\;\;\seql{$(+,\xi_3,I_2)$}{\cdots\;\;\xi_3i_2\vdash\Lambda_5\;\;\cdots}{\vdash\xi_2i_1\star J_2,\Lambda_4}\cdots\;\;}{\xi_2i_1\vdash\Lambda_3}\;\;\cdots}{\vdash \xi_1\star J_1,\Lambda_2}\;\;\cdots}{\xi_1\vdash\Lambda_1}$$

\smallskip
Note that the definition allows infinite designs, because the negative rule allows for infinite branching.
But we also allow vertical infinity, i.e., infinite depth, and in particular we allow for recursive definitions
of designs, as in the following fundamental example. The {\em fax} ${\it Fax}_{\xi,\xi'}$ based on $\xi\vdash \xi'$ is the infinite design which is recursively specified as follows:

$$\begin{array}{ccccc}
{\it Fax}_{\xi,\xi'} && = && \seql{$(-,\xi,{\cal P}_f(\omega))$}{\cdots\quad\seql{$(+,\xi',J_1)$}{\cdots\quad{\it Fax}_{\xi'j_1,\xi j_1}\quad\cdots}{\vdash\xi\star J_1,\xi'}\quad\cdots}{\xi\vdash\xi'}
\end{array}$$

\noindent
 As its name suggests, the fax is a design for copying, which is the locative variant of identity, as it maps data to the \guil{same} data, but copied elsewhere. It is indeed folklore from the work on game semantics (see e.g. \cite{AbraMcCus99})  -- and actually from the earlier work of \cite{BerryCurien82} -- that
computations are just mosaics of small bricks consisting of such copycat 
programs.
Partial (finite or infinite) versions of the fax are obtained by deciding to place some ${\cal N}$'s instead of having systematically
${\cal P}_f(\omega)$ everywhere.  How to define infinite designs of this kind and solve such recursive equations precisely is the matter of exercise \ref{designs-recursive}. Other simple examples of designs are given in definition \ref{dai-skunk-ram-dir}.

\begin{remark} \label{implicit-weakening}
Note that the three rules embody implicit potential weakening. Recall that weakening consists in deriving $\Gamma,\Gamma'\vdash\Delta,\Delta'$ from
$\Gamma\vdash\Delta$ (for arbitrary $\Gamma'$ and $\Delta'$). Indeed, in the Daimon rule, $\Lambda$ is arbitrary, and in the two other rules we only require that $\Lambda$ contains the union of the $\Lambda_i$'s (resp. the $\Lambda_J$'s) (see also Remark \ref{affine}).
\end{remark}

\begin{remark} \label{libre-lie}
It is immediate that the addresses appearing in a design based on $\vdash \Lambda$ (resp. on $\zeta\vdash\Lambda$) are extensions $\xi\xi'$ of addresses $\xi$ of $\Lambda$ (resp. $\set{\zeta}\union\Lambda$).
Also, it is easy to see that the property of well-formedness is forced upon designs whose  base has the form  $\vdash \xi$ or $\xi\vdash$, as well as the following one (the parity of an address refers to whether its length is even or odd):
\begin{itemize}
\item All addresses in each base on the right of $\;\vdash$ have the same parity, and then the formula on the left if any has the opposite parity.
\end{itemize}
Girard requires this propertiy explicitly in the definition of a base. It does not appear to play an essential role, though.
\end{remark}

\Proofitem{\mbox{\bf Two flavors of designs.}} 
Girard also gives an alternative definition of designs, as strategies.
Let us briefly recall the dialogue game interpretation of proofs (cf. part I, section 2). In this interpretation, proofs are considered as strategies for a Player playing against an Opponent.
The Player plays inference rules, and the opponent chooses  one of the premises in order to challenge the Player to disclose the last inference rule used to establish this premise, etc... . In a positive rule,  Player chooses a $\xi$ and an $I$. (Going back to the setting of section \ref{ludics}, he chooses which formula ($\xi$) to decompose, and which synthetic rule ($I$) to apply.)
But in a negative rule $(-,\xi,{\cal N})$, Player does not choose the $\xi$ (cf. Remark \ref{neg-pas-choix}). Moreover, we can equivalently formulate the information conveyed by ${\cal N}$ by adding a premise $\Omega$ (``undefined") for each   $I\nin{\cal N}$: intuitively, the non-presence of $I$ in ${\cal N}$ is not a choice of Player, but rather a deficiency of her  strategy: she has not enough information to answer an \guil{attack} of Opponent that would interrogate this branch of the tree reduced to $\Omega$. These remarks suggest a formulation where the negative rule disappears, or rather is merged with the positive rule:

$$\seql{$(+,\xi,I)$}{\overbrace{\cdots\quad\overbrace{\cdots\quad\vdash\xi i\star J,\Lambda_{i,J}\quad\cdots}^{J\in{\cal P}_f(\omega)}\quad\cdots}^{i\in I}}{\vdash\xi,\Lambda}$$

\noindent
where $\Lambda_{i_1,J_1}$ and $\Lambda_{i_2,J_2}$ are disjoint as soon as $i_1\neq i_2$. 
And we need to introduce $\Omega$:
$$\seql{$\Omega$}{}{\vdash\Lambda}$$

\noindent
This axiom is used for each $J'\in{\cal P}_f(\omega)\setminus{\cal N}$. 
Note that this change of point of view leads us to add a new tree that did not exist in the previous definition, namely the tree reduced to the $\Omega$ axiom. Girard calls it the {\em partial design}.

\begin{remark} \label{union-finie-infinis}
Note that in the above formulation of the rule, there are infinitely many premises, that are split 
into a finite number of infinite sets of premises, one for each $i\in I$.  The syntax that we shall introduce below respects this layering.
\end{remark}

\smallskip
Under this presentation of designs, an Opponent's move consists in picking an $i$ (or $\xi i$), and a $J$.
We thus use names $(-,\zeta,J)$ to denote Opponent's moves after a rule $(+,\xi,I) $, with $\zeta=\xi i$ for some $i\in I$.
Here is how the generic example above gets reformulated:

$$\cdots\; \seql{$\Omega$}{}{\vdash\xi_1\star J'_1,\Lambda'_2}\;\cdots\;\seql{$(+,\xi_2,I_1)$}{\cdots\; \seql{$\Omega$}{}{\vdash\xi_2i_1\star J'_2,\Lambda'_4}\;\cdots\;\seql{$(+,\xi_3,I_2)$}{\cdots}{\vdash\xi_2i_1\star J_2,\Lambda_4}\;\cdots}{\vdash \xi_1\star J_1,\Lambda_2}\;\cdots$$

\noindent
Finally, we can get rid of the  \guil{abstract sequents}, and we obtain a tree (resp. a forest) out of a positive (resp. negative) design, whose nodes are labelled by
alternating moves $(+,\xi,I)$ and $(-,\zeta,J)$, which are called  positive and negative {\em actions}, respectively:

$$\cdots\; \seql{$\Omega$}{}{(-,\xi_1, J'_1)}\;\cdots\;\seql{$(+,\xi_2,I_1)$}{\cdots\;\seql{$\Omega$}{}{(-,\xi_2i_1, J'_2)}\;\cdots\;\seql{$(+,\xi_3,I_2)$}{\cdots}{(-,\xi_2i_1, J_2)}\;\cdots}{(-,\xi_1,J_1)}\;\cdots$$

\noindent
By abuse of language (especially for $\Omega$), $\Omega$ and $\demon$ can also be called positive actions.
We say that $(+,\xi,I)$ (resp. $(-,\zeta,J)$) is focalized in $\xi$ (resp. $\zeta$). The {\em opposite} (or dual) of an action $(+,\xi,I)$ (resp. $(-,\xi,I)$) is the action $(-,\xi,I)$ (resp. $(+,\xi,I)$).

\smallskip
Girard describes the tree or forest 
of actions as its set of branches, or
{\em chronicles}, which are alternating
sequences of actions (see Remark \ref{design-chronicles}). A chronicle ending with a negative (resp. positive) action is called negative (resp. positive).  Here, we choose a slightly different approach, and we present a {\em syntax} for designs. More precisely, we first present {\em pseudo-designs}, or raw designs, and we use a typing system (or, rather, a sorting system) to single out the designs as the well-typed pseudo-designs.

\medskip\noindent
Here is the syntax:

$$\begin{array}{lllll}
\mbox{Positive pseudo-designs} &&&&
\phi::= \Omega\Alt\demon\Alt (+,\xi,I)\cdot\setc{\psi_{\xi i}}{i\in I}\\
\mbox{Negative pseudo-designs} &&&&
\psi_\zeta::=\setc{(-,\zeta,J)\phi_J}{J\in{\cal P}_f(\omega)}
\end{array}$$

\noindent
We shall use the notation $\psi_{\zeta,J}$ for $\phi_J$.  We remark:
\begin{itemize}
\item the r\^ole of $I$ that commands the cardinality of the finite set of  the $\psi_{\xi i}$'s;
\item the uniform indexation by the set of all finite parts of $\omega$ of the set of trees of a pseudo-design $\psi_\zeta$;
\item the r\^ole of the index $\zeta$ in $\psi_\zeta$, that commands the shape of its initial negative actions.
\end{itemize}

\noindent
The syntax makes it clear that a positive design is a tree the root of which (if different from $\Omega$ and $\demon$)
is labelled by a positive action $(+,\xi,I)$ and has branches indexed by $i\in I$ and $J\in {\cal P}_f(\omega)$. The corresponding subtrees are grouped in a layered way (cf. Remark \ref{union-finie-infinis}): for each $i$, the subtrees intexed by $i,J$, for $J$ varying over ${\cal P}_f(\omega)$ form a negative design
$\psi_{\xi i}$, each of whose trees has a root labelled with the corresponding negative action $(-,\xi i,J)$.
Note that each negative root $(-,\xi i,J)$ has a unique son: designs are strategies, i.e., Player's answers to Opponent's moves are unique.

\medskip
The following definition collects a few useful designs expressed in our syntax.

\begin{definition} \label{dai-skunk-ram-dir} We set:

$$\begin{array}{l}
{\it Dai} = \demon\\
{\it Dai}^-_\xi = \setc{(-,\xi,J)\demon}{J\in{\cal P}_f(\omega)}\\
{\it Skunk}_\xi = \setc{(-,\xi,J)\Omega}{J\in{\cal P}_f(\omega)}\\
{\it Skunk}^+_{(\xi,I)}= (+,\xi,I)\cdot\setc{\setc{(-,\xi i,J)\Omega}{J\in{\cal P}_f(\omega)}}{i\in I}\\
{\it Ram}_{(\xi,I)}=(+,\xi,I)\cdot\setc{\setc{(-,\xi i,J)\demon}{J\in{\cal P}_f(\omega)}}{i\in I}\\
{\it Dir}_{\cal N}=\setc{(-,\epsilon,I)\demon}{I\in{\cal N}}\union
\setc{(-,\epsilon,I)\Omega}{I\nin{\cal N}}
\end{array}$$
\end{definition}


\noindent
The skunk is an animal that stinks and has therefore degree zero of  sociability, where sociability is measured by the capacity for compromise, expressed by $\demon$ in ludics (see Exercise \ref{skunk-daimon}).

\medskip 
Not every  term of this syntax corresponds to a design, whence our terminology of pseudo-design.
We can recover those pseudo-designs that come from designs as the \guil{correctly typed} ones. More precisely, designs can be defined 
either as typing proofs of pseudo-designs (our first definition earlier in the section), or as
 typable pseudo-designs. Girard uses the French words \guil{dessin} and \guil{dessein} to name these
 two flavours, respectively. This is reminiscent of the distinction between typing \guil{\`a la Church} and typing \guil{\`a la Curry}, respectively.
Here is the \guil{type system}:

$$\begin{array}{lll}
\seq{(\vdash\Lambda\mbox{ well formed})}{\Omega:(\vdash \Lambda)} & \quad\quad & \seq{(\vdash\Lambda\mbox{ well formed})}{\demon:(\vdash\Lambda})
\end{array}$$

$$\seq{\cdots\;\psi_{\xi i}:(\xi i\vdash \Lambda_{i})\;\cdots\quad\left(\begin{array}{l}
i\in I\\
\qqs{i}{\Lambda_{i}\inc\Lambda}\\\qqs{i_1,i_2}{(i_1\neq i_2 \Rightarrow \Lambda_{i_1}\inter\Lambda_{i_2}=\emptyset)}\\
\vdash\xi,\Lambda\mbox{ well formed}
\end{array}\right)}
{(+,\xi,I)\cdot\setc{\psi_{\xi i}}{i\in I}:(\vdash\xi,\Lambda)}$$

$$\seq{\cdots\;\phi_J:(\vdash \zeta \star J,\Lambda_{J})\;\cdots\quad\left(\begin{array}{l}
J\in{\cal P}_f(\omega)\\
\qqs{J}{\Lambda_{J}\inc\Lambda}\\
\zeta\vdash\Lambda\mbox{ well formed}
\end{array}\right)}
{\setc{(-,\zeta,J)\phi_J}{J\in {\cal P}_f(\omega)}:(\zeta\vdash\Lambda)}$$

\smallskip


\noindent
We also write
$\Psi:\set{\ldots,(\zeta_i\vdash\Lambda_i),\ldots}$  if 
 $\Psi=\set{\psi_{\zeta_1},\ldots,\psi_{\zeta_n}}$ and $\psi_{\zeta_i}:(\zeta_i\vdash\Lambda_i)$ for all $i$
 (finite number of bases).


\begin{definition}
A design is a  typable pseudo-design $\neq\Omega$. The pseudo-design  $\Omega$, which is obviously typable ($\Omega:(\vdash\Lambda)$, for any $\Lambda$), also denoted by ${\it Fid}$, is called the partial design.
\end{definition}

\medskip
Conversely, the following algorithm takes as input a finite positive pseudo-design $\phi$ and returns a base $\vdash\Lambda$ such that $\phi:(\vdash\Lambda)$ if it is a design and otherwise returns ${\it FAIL}$.
Actually, the inductive load requires taking as
arguments $\phi$ together with a (well-formed) base $\vdash\Gamma$. Initially, we take $\Gamma$ empty. The algorithm is described by a formal system using the following notation:
$$\phi:(\vdash\Gamma,?)\longrightarrow\phi:(\vdash\Lambda)$$
which reads as follows:
the algorithm, when given $\phi$ and $\Gamma$ as inputs, finds $\Lambda$ such that   $\vdash\Lambda$ (cf. Remark \ref{libre-lie}), $\Gamma\inc\Lambda$, and $\phi:(\vdash\Lambda)$. The failure case is written as $\phi:(\vdash\Gamma,?)\longrightarrow{\it FAIL}$.
The induction is on $\phi$. If the design is infinite, we apply the algorithm to finite approximations -- we can obviously not decide whether an infinite pseudo-design is a design.

$$\begin{array}{lll}
\seq{}{\Omega:(\vdash\Gamma,?)\longrightarrow\Omega:(\vdash \Gamma)} && 
\seq{}{\demon:(\vdash\Gamma,?)\longrightarrow\demon:(\vdash\Gamma)}
\end{array}$$

\smallskip\noindent
In the following rules, the notation $\Psi_{\xi i,J}$ stands for $\phi_J$ where
$\Psi=\setc{\psi_{\xi i}}{i\in I}$ and $\psi_{\xi i}=\setc{(-,\zeta,J)\phi_J}{J\in{\cal P}_f(\omega)}$.

$$\seq{\cdots\;\Psi_{\xi i,J}:(\vdash \xi i\star J,?)\longrightarrow\Psi_{\xi i,J}:(\vdash \xi i\star J,\Lambda_{i,J})\;\cdots\quad\left(\begin{array}{l}
i\in I, J\in{\cal P}_f(\omega)\\
\qqs{i_1,i_2}{(i_1\neq i_2 \Rightarrow \Lambda_{i_1,J_1}\inter\Lambda_{i_2,J_2}=\emptyset)}\\
\vdash \xi,\Gamma\setminus\set{\xi}\Union_{i\in I, J\in{\cal P}_f(\omega)}\Lambda_{i,J}\mbox{ well formed}
\end{array}\right)}
{(+,\xi,I)\cdot\Psi:(\vdash\Gamma,?)\longrightarrow(+,\xi,I)\cdot\Psi:(\vdash\xi,\Gamma\setminus\set{\xi}\Union_{i\in I, J\in{\cal P}_f(\omega)}\Lambda_{i,J})}$$

$$\seq{\Psi_{\xi i,J}:(\vdash \xi i\star J,?)\longrightarrow{\it FAIL}}
{(+,\xi,I)\cdot\Psi:(\vdash\Gamma,?)\longrightarrow{\it FAIL}}$$

$$\seq{\begin{array}{l}\Psi_{\xi i_1,J_1}:(\vdash \xi i_1\star J_1,?)\longrightarrow\Psi_{\xi i_1,J_1}:(\vdash \xi i_1\star J_1,\Lambda_{i_1,J_1})
\\
\Psi_{\xi i_2,J_2}:(\vdash \xi i_2\star J_2,?)\longrightarrow\Psi_{\xi i_2,J_2}:(\vdash \xi i_2\star J_2,\Lambda_{i_2,J_2})
\end{array}\quad(i_1\neq i_2 \mbox{ and } \Lambda_{i_1,J_1}\inter\Lambda_{i_2,J_2}\neq\emptyset)}
{(+,\xi,I)\cdot\Psi:(\vdash\Gamma,?)\longrightarrow{\it FAIL}}$$

$$\seq{\cdots\;\Psi_{\xi i,J}:(\vdash \xi i\star J,?)\longrightarrow\Psi_{\xi i,J}:(\vdash \xi i\star J,\Lambda_{i,J})\;\cdots\quad
(\vdash\xi,\Gamma\setminus\set{\xi}\Union_{i\in I, J\in{\cal P}_f(\omega)}\Lambda_{i,J}\mbox{ not well formed})}
{(+,\xi,I)\cdot\Psi:(\vdash\Gamma,?)\longrightarrow{\it FAIL}}$$


\noindent
The first rule of failure, which has a unique premise (for a given $i\in I$ and a given $J$), expresses failure propagation. The second rule  says  that failure occurs when linearity (actually, affinity)
is violated. The third rule controls the well-formedness of bases.
The following statement is easy to prove.

\begin{proposition} The above algorithm, with $\phi,\Gamma$ as arguments, terminates if and only if there exists $\Lambda'$ such that $\phi:(\vdash\Lambda')$ and $\Gamma\inc\Lambda'$. The base $\vdash  \Lambda$ returned by the algorithm is the smallest base that satisfies these conditions.
\end{proposition}

\Proofhint That the algorithm returns the minimum base follows from the fact that
it collects unions of bases, which are themselves least upper bounds. \qed

\begin{remark}
The relation between the two flavours of designs (dessins and desseins, or typing proofs and typable pseudo-designs) is of the same nature as the relation between proofs in natural deduction expressed in terms of sequents
and proofs in natural deduction expressed as trees of formulas with marked hypotheses, or as the relation between linear logic proofs  in the sequent calculus formulation (cf. part I, section 2)  and proof nets (cf. section \ref{proof-nets}). While proof nets allow us to identify many sequent calculus proofs,
here, as we have just seen, the two points of view (explicit sequents or not) happen to be essentially the same, up to weakening. This is due  to the
focalization, which already eliminated many sequent calculus proofs!
\end{remark}

\begin{remark} \label{design-chronicles}
We can define an injection ${}^\bullet$ mapping pseudo-designs to sets of chronicles ending with a positive action as follows:

$$\begin{array}{lll}
\Omega^\bullet = \emptyset & & ((+,\xi,I)\cdot\setc{\psi_{\xi i}}{i\in I})^\bullet=\set{\epsilon}\Union_{i\in I} \setc{(+,\xi,I)r}{r\in \psi_{\xi i}^\bullet}\\
\demon^\bullet = \set{\demon} & &
\setc{(-,\zeta,J)\phi_J}{J\in{\cal P}_f(\omega)}^\bullet=\Union_{J\in{\cal P}_f(\omega)}\setc{(-,\zeta,J)r}{r\in\phi_J^\bullet}
\end{array}$$

\noindent Note that the $\Omega$'s disappear during this transformation. In \cite{Gir01}, the designs-desseins are defined as sets of chronicles, and properties characterizing the image of the set of designs by the function ${}^\bullet$ are given. These properties include:

\Proofitem{\bullet}
{\em coherence} : the maximal prefix of any two chronicles of the design which are not prefix of each other ends with a positive action;

\Proofitem{\bullet}
{\em focalization}:  a negative action which is not at the root must  be of the form $(-,\xi i,J)$ and its father must be of the form $(+,\xi,I)$, with $i\in I$;

\Proofitem{\bullet}
{\em subaddress}: a positive action $(+,\xi,I)$ is either such that $\xi$ is an address on the positive side of the base, or there exist $\xi',i$, and $J$ such that $\xi=\xi' i$, $i\in J$, and  $(-,\xi',J)$ occurs on the chronicle between $(+,\xi,I)$ and the root.
\end{remark}

\smallskip
Designs have an implicit pointer structure: for each action whose focus is some $\xi i$, one looks for the action of focus $\xi$ below it in the forest or tree representation of the design, if any, and we say that the former {\em points} to the latter, or is {\em bound} to the latter. Otherwise, we say that the action of focus $\xi i$ occurs {\em free} in the design.
 It is easily seen that for well-typed designs
the two actions must have opposite signs, and that a free action is necessarily positive and that its focus belongs to the right hand side of the base of the design (cf. Remark \ref{libre-lie}). Moreover, negative actions are always bound and always point to the immediately preceding positive action.
These are  typical features of Hyland and Ong (HO)  games \cite{HO2000} (see \cite{FagHyl02} for precise comparisons). 
In  \cite{CH98,Cu98}, a general formalism of such (untyped) HO style strategies over a given alphabet of moves,  called {\em abstract B\"ohm trees}, has been developped. Opponent's moves are letters of the alphabet, while Player's moves are pairs noted $\ppm{a}{\kappa}$, where $a$ is a letter of the alphabet and where $\kappa$ is either the symbol $\_$ that indicates that the move is free, or a natural number  that counts the number of Opponent's moves  encountered before reaching the binder.
The designs are  abstract B\"ohm trees constructed on the alphabet consisting of all the pairs $(i,I)$ of a natural number and a finite set of natural numbers. 
The general shape is as follows (rotating the tree clockwise by 90 degrees):

$$\begin{branch}\vdots\\
(j_1,J'_1) \Omega\\
\vdots\\
(j_1,J_1)\ppm{(j_2,I_1)}{\kappa_2}\begin{branch}\vdots\\
(i_1,J'_2)\Omega\\
\vdots\\
(i_1,J_2)\ppm{(j_3,I_2)}{\kappa_3}\begin{branch}\vdots\end{branch}\\
\vdots
\end{branch}\\
\vdots
\end{branch}$$ 

\bigskip
\noindent
The difference with the (raw) designs is that we use relative addressses rather than absolute ones.
The pointers allow us to reconstruct the full addresses.  Here is how the fax looks like, first,
as a design-dessein:

$$\begin{branch}
\vdots\\
(-,\xi,J_1)(+,\xi',J_1)\begin{branch}\vdots\\
(-,\xi' j_1,J_2)(+,\xi j_1,J_2)\begin{branch}\vdots\\
(-,\xi j_1j_2,J_3)(+,\xi 'j_1j_2,J_3)\begin{branch}\vdots
\end{branch}\\
\vdots
\end{branch}\\
\vdots\\
\end{branch}\\
\vdots
\end{branch}$$

\noindent 
and then as  abstract B\"ohm tree
(where $\xi,\xi'$ end with $i,i'$, respectively):

$$\begin{branch}
\vdots\\
(i,J_1)\ppm{(i',J_1)}{\_}\begin{branch}\vdots\\
(j_1,J_2)\ppm{(j_1,J_2)}{1}\begin{branch}\vdots\\
(j_2,J_3)\ppm{(j_2,J_3)}{1}\begin{branch}\vdots
\end{branch}\\
\vdots
\end{branch}\\
\vdots\\
\end{branch}\\
\vdots
\end{branch}$$

\noindent
The pointer 1 in, say,  $\ppm{(j_1,J_2)}{1}$, formalizes the fact that  the positive action 
$(+,\xi j_1,J_2)$ is not bound
to the negative action that immediately precedes it, but, one level up, to $(-,\xi,J_1)$: Girard calls this subfocalization.

\Proofitem{\mbox{\bf Designs and $\lbd$-calculus.}}
If we restrict both designs and (a variant of) $\lbd$-calculus, then we
can get a bijective correspondence, as we show now. The interest is twofold. First, this correspondence allows us to connect designs to the well-established tradition of the $\lbd$-calculus. 
But also, it has suggested us to extend the correspondence to all designs, giving rise to a term language for designs.
We start with the restricted correspondence, which applies to slices, defined next.

\begin{definition}[Slice] \label{slice}
A  slice is a design in which each application of the rule
$(-,\xi,{\cal N})$ is such that $\cal N$ is a singleton. In terms of designs as sets of chronicles, this rule says  that if the design contains two chronicles $r(-,\xi,I_1)$ and $r(-,\xi,I_2)$, then $I_1=I_2$. 
\end{definition}

\noindent
(It is easy to check that slices arise from purely multiplicative proofs.)
We introduce variables and proof terms for slices simultaneously, as follows:

$$\begin{array}{lll}
\seq{}{\Omega:(\vdash \Lambda)} & \quad\quad & \seq{}{\demon:(\vdash\Lambda})
\end{array}$$

$$\seq{P:(\vdash \setc{x_j:\zeta j}{j\in J},\Lambda')\quad(\Lambda'\inc\Lambda)}
{\lbd \setc{x_j}{j\in J}.P:(\zeta\vdash\Lambda)}$$

$$\seq{\cdots\;M_i:(\xi i\vdash \Lambda_{i})\;\cdots\quad\left(\begin{array}{l}
i\in I\\
\qqs{i}{\Lambda_{i}\inc\Lambda}\\\qqs{i_1,i_2}{(i_1\neq i_2 \Rightarrow \Lambda_{i_1}\inter\Lambda_{i_2}=\emptyset)}\\
x \mbox{ fresh}\end{array}\right)}
{x\setc{M_i}{i\in I}:(\vdash x:\xi,\Lambda)}$$

\noindent 
In the first (resp. second) rule, the cardinal of $J$ (resp. $I$) gives the number of head $\lbd$'s (resp. the number of arguments of the variable).
The bases $\xi\vdash\Lambda$ and $\vdash\Lambda$ are now such that $\Lambda$ consists of a set of variable declarations of the form $x:\zeta$.
Let us place the $\lbd$-calculus style syntax that we just introduced (on the left below) in perspective with the syntax of normal forms of the $\lbd$-calculus (on the right below):

$$\begin{array}{lllll}
M::= \lbd \setc{x_j}{j\in J}.P &&&& M::= \lbd x_1\ldots x_m.P\\
P::= x\setc{M_i}{i\in I}\Alt\Omega\Alt\demon &&&& P::= xM_1\ldots M_n
\end{array}$$

\noindent
The difference lies in the fact that we have freed ourselves from the sequential order of application of the ordinary $\lbd$-calculus, and that we now have explicit addresses for the arguments of the application: $i$  is the (relative) address of $M_i$.

\begin{remark}
Combining the bijection from slices to terms just given with the representation of designs as HO style trees with pointers given above, what we have obtained
is a correspondence of the same nature as the bijection between $\lambda$-terms and $\lambda$-terms in De Bruijn notation (see e.g. \cite{ACCL92}). More precisely, it is exactly a bijection of this kind, replacing De Bruijn indices by  other indices that we have called B\"ohm indices in  \cite{Cu98} (and that have appeared in previous works of Danos-Regnier and Huet), which are pairs of indices of the form $\ppm{i}{j}$ where
$j$ counts the number of $\lbd$'s separating the variable from its binder and where $i$ serves to identify which among the variables collectively bound by the binder it is. For example, the B\"ohm indices of $x_1$ and $x_2$ in
$\lbd\set{x_1,x_2}.x_1\set{\lbd\set{y_1}.x_2\set{}}$ are $\ppm{1}{0}$ and
$\ppm{2}{1}$, respectively.
\end{remark}

\begin{remark} \label{affine}
It is easily checked that the typable terms are exactly the affine terms, i.e., those in which any variable occurs at most once. 
\end{remark}

\medskip
We next move on and give terms for {\em arbitrary designs}.

$$\begin{array}{lll}
\seq{}{\Omega:(\vdash \Lambda)} & \quad\quad & \seq{}{\demon:(\vdash\Lambda})
\end{array}$$

$$\seq{\cdots\;M_i:(\xi i\vdash \Lambda_{i})\;\cdots\quad\left(\begin{array}{l}
i\in I\\
\qqs{i}{\Lambda_{i}\inc\Lambda}\\\qqs{i_1,i_2}{(i_1\neq i_2 \implies \Lambda_{i_1}\inter\Lambda_{i_2}=\emptyset)}\\ x \mbox{ fresh}\end{array}\right)}
{(x.I)\cdot\setc{M_i}{i\in I}:(\vdash x:\xi,\Lambda)}$$

$$\seq{\cdots\;P_J:(\vdash \setc{x_j:\zeta j}{j\in J},\Lambda_{J})\;\cdots\quad\left(\begin{array}{l}
J\in{\cal P}_f(\omega)\\
\qqs{J}{\Lambda_{J}\inc\Lambda}\end{array}\right)}
{\setc{J=\lbd\setc{x_j}{j\in J}.P_J}{J\in{\cal P}_f(\omega)}:(\zeta\vdash\Lambda)}$$

\medskip
\noindent
Forgetting now about typing rules, we have the following (untyped) syntax:

$$\begin{array}{l}
M::=  \setc{J=\lbd\setc{x_j}{j\in J}.P_J}{J\in{\cal P}_f(\omega)}\\
P::=(x\cdot I)\setc{M_i}{i\in I}\Alt \Omega \Alt \demon
\end{array}$$

\noindent
Since we now have two syntaxes for (pseudo-)designs, we shall refer to the two syntaxes as the abstract syntax and the concrete syntax, respectively (in the order in which they have appeared in the paper).
As an illustration, here is the fax (recursively) expressed in concrete syntax:

$${\it Fax}_{\xi,x:\xi'}\quad=\quad\setc{I=\lbd\setc{y_i}{i\in I}.(x\cdot I)\setc{{\it Fax}_{\xi' i,y_i:\xi i}}{i\in I}}{I\in{\cal P}_f(\omega)}\;.$$

\noindent In this form, the fax appears as the variable $x$, in all its possible (hereditarily) $\eta$-expansions. Note the remarkable use of the additive flavour of designs: the fax is ready for {\em any}  ramification $I$, i.e. for any possible $\eta$-expansion (see Exercise \ref{fax-renaming}).

\bigskip\noindent
\Proofitem{\mbox{\bf Ordering designs.}} Two rather natural partial orders can be defined on designs:

\Proofitem{\bullet} The usual partial information ordering on trees: to increase information, one replaces an $\Omega$ with a tree $\neq\Omega$ (in the first representation of designs, this amounts to extend a set ${\cal N}$ in a negative rule). This is reminiscent of the ordering between stable functions defined by the inclusion of their traces (see e.g. \cite[chapter 12]{AmaCur}).

\Proofitem{\bullet} The second ordering $\sqinc$ is obtained by adding a second rule for increasing: replace a subtree by a $\demon$. We shall see that this ordering characterizes the observational order between designs, so we shall call it the observational ordering. It is reminiscent of the pointwise ordering between Scott continuous functions. This ordering is defined formally as follows:

\smallskip

$$\begin{array}{ccc}
\seq{}{\Omega\sqinc\phi} && \seq{}{\phi\sqinc\demon}
\\\\\seq{\cdots\quad\psi_{\xi i}\sqinc\psi'_{\xi i}\quad\cdots\quad\quad (i\in I)}{(+,\xi,I)\cdot\setc{\psi_{\xi i}}{i\in I}\sqinc(+,\xi,I)\cdot\setc{\psi'_{\xi i}}{i\in I}} &&
\seq{\cdots\quad\phi_J\sqinc\phi'_J\quad\cdots\quad\quad J\in{\cal P}_f(\omega)}
{\set{\ldots,(-,\zeta,J)\phi_J,\ldots}\sqinc\set{\ldots,(-,\zeta,J)\phi'_J,\ldots}}
\end{array}$$

$$\begin{array}{ccc}
\seq{}{\phi\sqinc\phi} && \seq{\phi_1\sqinc\phi_2\quad\phi_2\sqinc\phi_3}{\phi_1\sqinc\phi_3}
\end{array}$$

\noindent 
We write $\phi_1\leq^R\phi_2$ (resp. $\phi_1\leq^L\phi_2$) if $\phi_1\sqinc\phi_2$ has been proved without using the axiom $\phi\sqinc\demon$ (resp. $\Omega\sqinc\phi$).  Thus, $\phi_1\leq^R\phi_2$
is the stable ordering, which we also denote as $\phi_1\inc\phi_2$ (see Exercise \ref{orders-chronicles}).

\begin{remark} We have the following decomposition property: if
$\phi_1\sqinc\phi_2$, then there exists a $\phi$ such that $\phi_1\leq^L\phi\leq^R\phi_2$, obtained by first cutting some subtrees and replacing them by a $\demon$, and then by adding some subtrees at some $\Omega$'s. Note that this is a quite natural way of proceeding, since one might do useless work otherwise, by adding subtrees using axiom $\Omega\sqinc\phi$ that could then be removed while using axiom $\phi\sqinc\Omega$.  The decomposition is not unique, as we have both
$\Omega\leq^L\Omega\leq^R\demon$ and $\Omega\leq^L\demon\leq^R\demon$ (see Exercise \ref{designs-Laird}).  Remarkably enough, similar decompositions have been discovered and exploited elsewhere:
\begin{itemize}
\item by Winskel in his analysis 
 of Berry's stable ordering and bi-domains (see \cite{CPW97}), and 
 \item by Laird in his extensional account of the model of sequential algorithms \cite{Laird2003,CurSA2004}.
 \end{itemize}


\end{remark}

\begin{exercise}
Complete the characterization of designs as sets of chronicles (cf. Remark \ref{design-chronicles}).
(Hint: The only missing conditions are the ones ensuring that the designs are affine.)
\end{exercise}

\begin{exercise} Is there a sense in which the concrete syntax and the abstract syntax of pseudo-designs  or of designs are in bijective correspondence? (Hints:  Pseudo-designs in abstract syntax are more permissive, e.g., they do not exclude the possibility of a move $(+,\xi i,J)$ ``pointing" to a move $(+,\xi, I)$.
On the other hand, ``raw" terms in the concrete syntax are actually all ``typable", in a system where the affinity constraint is relaxed.)
\end{exercise}

\begin{exercise} \label{orders-chronicles}
Show that, on designs as sets of chronicles (cf. Remark \ref{design-chronicles}), the stable ordering is set inclusion, and that the extensional ordering can be characterized by the following property:  $\phi_1\sqinc\phi_2$ if and only if whenever  $r_1=r\:(-,\xi,I)r'_1\in\phi_1$ and 
$r_2=r\:(-,\xi,I)r'_2\in\phi_2$ diverge after some action $(-,\xi,I)$, then either $r'_1=\Omega$ or $r'_2=\demon$.
\end{exercise}

\begin{exercise} \label{designs-recursive}
We define formally (infinite-depth) designs as ideals (i.e., downwards closed directed subsets of
designs) with respect to $\inc$. Show that the equation defining the fax admits a least fixed point which is such an ideal.
\end{exercise}

\begin{exercise} \label{designs-Laird}
Let $\leq$ be the intersection of the partial orders $\leq^L$ and $\leq^R$. Show the following more precise formulation of
the decomposition property:   if
$\phi_1\sqinc\phi_2$, then there exist  $\phi_{\it min}$ and $\phi_{\it max}$ such that:
$$\qqs{\phi} {((\phi_1\leq^L\phi\leq^R\phi_2) \Leftrightarrow \phi_{\it min}\leq\phi\leq \phi_{\it max})}\;.$$ 
\end{exercise}

\section{Normalization of designs} \label{normalization}

Normalization  is a general machinery that applies to all abstract B\"ohm trees, hence to designs in particular. It can be described in various equivalent ways, see \cite{CH98}.
In the affine case, the execution is simpler (no  copying involved), so we shall describe here (in three disguises) an abstract machine for designs taking the affine nature of designs into account. 
The first two presentations of the normalization engine are based on the abstract and concrete syntax for designs, respectively, and use the framework of environment machines, whose states have two components: the code to be run, and the environment or context or counter-design. 
The third version is more visual, and features a simple token game. Its main advantage is that it
makes the very useful notion of the part of a design visited during normalization explicit.

\medskip  We embark on the first description, for which
 we shall need the following terminology.

\begin{notation} \label{grand-psi}
We denote by $\Psi$ a finite set
$\set{\psi_{\zeta_1},\ldots,\psi_{\zeta_n}}$
of negative pseudo-designs, with the $\zeta_i$'s pairwise disjoint.
We say that $\Psi$ {\em accepts} (resp. does not accept) an address $\zeta$ if $\zeta\in\set{\zeta_1,\ldots,\zeta_n}$
(resp. $\zeta\nin\set{\zeta_1,\ldots,\zeta_n}$). If $\Psi$ accepts $\zeta$, i.e., $\zeta=\zeta_j$ for some $j$, we  use the notation  $\Psi_\zeta$ for $\psi_{\zeta_j}$. 
We also note simply $\psi$ for $\set{\psi}$. Finally, if $\Psi$ accepts $\zeta$, we note $\Psi\setminus\zeta=\Psi\setminus\set{\psi_\zeta}$.
\end{notation}

We start with a design $\phi:(\vdash\xi,\Lambda_1)$ to be normalized against a design
$\psi:(\xi\vdash\Lambda_2)$ (with $\Lambda_1$ and $\Lambda_2$ pairwise disjoint), which we call its {\em counter-design}. The normalization should produce a design based on $\vdash\Lambda_1,\Lambda_2$ (think of a cut rule). 

The machine has states of the form $\coupe{\phi}{\Psi}$, and execution consists in applying the following state transformation, which we call the {\em weak reduction} rule (by contrast to strong reduction, introduced below):

$$\begin{array}{llll}
\mbox{(R)} & \coupe{(+,\xi,I)\cdot \Psi'}{\Psi}
 \longrightarrow 
\coupe{\Psi_{\xi,I}}{\Psi'\union(\Psi\setminus\xi)}
&&  (\Psi\mbox{ accepts }\xi)
\end{array}$$

\smallskip\noindent
Notice that we throw away all of $\Psi_\xi$ except for $\Psi_{\xi,I}$. In particular, the negative actions 
$(-,\xi,J)$ of $\Psi$ are not needed. That we can do this safely will be justified by Proposition 
\ref{pull-back-slice}.

\smallskip
The initial state is  $\coupe{\phi}{\set{\psi}}$.
It is important to keep $\Psi\setminus\xi$ in order to handle subfocalizations, as illustrated by the example below (where $\zeta=\xi i_1i$, for some $i\in I_1$):

$$\begin{array}{llllll}
\phi\quad =\quad\seq{\cdots\;\;\seq{\seq{\cdots}{(+,\zeta,J)}}{(-,\xi i_1,I_1)}\;\;\cdots\;\;\seq{\cdots}{(-,\xi i_2,I_2)}\;\;\cdots}{(+,\xi,I)} &&&&&
\psi\quad = \quad\cdots\;\;\seq{\seq{\cdots\;\;\seq{\seq{\cdots}{(+,\xi i_2,I_2)}}{(-,\zeta,J)}\;\;\cdots}{(+,\xi i_1,I_1)}}{(-,\xi,I)}\;\;\cdots
\end{array}$$

\smallskip\noindent or:
 
$$\begin{array}{ll}
\begin{array}{l}
\phi=(+,\xi,I)\cdot\Psi'\\
\quad\mbox{with }\Psi'=\set{\ldots,(-,\xi i_1,I_1)\phi_2,\ldots,(-,\xi i_2,I_2)\cdots,\ldots}\\
\quad\mbox{with }\phi_2=(+,\zeta,J)\cdot\Psi_2
\end{array} &
\begin{array}{l}
\psi=\set{\ldots,(-,\xi,I)\phi_1,\ldots}\\
\quad\mbox{with }\phi_1=(+,\xi i_1,I_1)\cdot\Psi_1\\
\quad\mbox{with }\Psi_1=\set{\ldots,(-,\zeta,J)\phi_3,\ldots}\\
\quad\mbox{with }\phi_3=(+,\xi i_2,I_2)\cdots
\end{array}
\end{array}$$

\noindent Execution goes as follows:

$$\begin{array}{lll}
\coupe{\phi}{\psi} & \longrightarrow & \coupe{\phi_1}{\Psi'\union(\psi\setminus\xi)}=
\coupe{\phi_1}{\Psi'}\\
& \longrightarrow & \coupe{\phi_2}{\Psi_1\union(\Psi'\setminus\xi i_1)}\\
& \longrightarrow &  \coupe{\phi_3}{\Psi_2\union(\Psi_1\setminus\xi)\union(\Psi'\setminus\xi i_1)}
\end{array}$$

\noindent
At state 
$\coupe{\phi_3}{\Psi_2\union(\Psi_1\setminus\xi)\union(\Psi'\setminus\xi i_1)}$, it is essential to have $\Psi'$ available, as the head action of $\phi_3$ corresponds to a head action of $\Psi'$ and not of $\Psi_2$, i.e.,
when normalization reaches $(+,\xi i_2,I_2)$, the corresponding negative action on the left is not found above $(+,\zeta,J)$, but above $(+,\xi,I)$.

\medskip
The machine stops when it reaches an $\Omega$ (which one should interpret as waiting for more information from the context), a $\demon$ (that indicates convergence), or a state 
$\coupe{(+,\xi_n,I_n)\cdot \Psi'_n}{\Psi_n}$ such that $\Psi_n$ does not accept $\xi_n$  (in $\lbd$-calculus terminology, this means reaching a head   variable). In all three cases, we have  found the root of the normal form of $\coupe{\phi}{\psi}$ (see also Remark \ref{var-tete-bien-typee}).

\smallskip
We can go on, and perform \guil{strong reduction}, by relaunching the machine
in order to get progressively, on demand, the chronicles of the
normal form, and not only the first positive action of the normal form. 
This can be formalized by indexing the state by the chronicle  $q$ of the normal form under exploration. The index remains invariant in the rule (R), which is thus reformulated as:

$$\begin{array}{llll}
\mbox{(R)} & \coupe{(+,\xi,I)\cdot \Psi'}{\Psi}_q
 \longrightarrow 
\coupe{\Psi_{\xi,I}}{\Psi'\union(\Psi\setminus\xi)}_q
&&  (\Psi \mbox{ accepts }\xi)
\end{array}$$

\noindent
Initially, the index  is the empty chronicle $\epsilon$. The chronicle is extended using the following rule ($S$ stands for strong reduction): 
$$\begin{array}{llll}
\mbox{(S)} & \coupe{(+,\xi,I)\cdot\Psi'}{\Psi}_q \longrightarrow \coupe{\Psi'_{\xi i,J}}{\Psi}_{q(+,\xi,I)(-,\xi i,J)} && (\Psi \mbox{ does not accept }\xi\mbox{ and }
i\in I)
\end{array}$$

\noindent 
Note that this rule is non-deterministic: the choice of $i,J$ is left to the Opponent of the normal form, he has to say which branch of the normal form he intends to explore.

The two other termination cases remain termination cases in this setting, and are formalized as follows: 

$$\begin{array}{l}
\coupe{\Omega}{\Psi}_q \longrightarrow\; !\,q\,\Omega \\
\coupe{\demon}{\Psi}_q \longrightarrow \;!\,q\,\demon
\end{array}$$

\noindent 
Read $!\,q\,\Omega$ (resp. $!\,q\,\demon$) as an output, saying that the chronicle $q\,\Omega$ (resp. $q\,\demon$) belongs to the normal form.  As for rule (S), we can read that $q(+,\xi,I)$ belongs to the normal form, but in addition we know that the normal form contains a chronicle of the form $q(+,\xi,I)(-,\xi i,J)\kappa_{i,J}$ for all $i\in I$ and $J$, and  may start an exploration in order to find any of these actions $\kappa_{i,J}$.

We denote the normal form of $\coupe{\phi}{\set{\psi}}$ by $\dl\phi,\set{\psi}\dr$, or simply $\dl\phi,\psi\dr$, and more generally, we denote the normal form of $\coupe{\phi}{\Psi}$ (see Definition \ref{designs-net}) by  $\dl\phi,\Psi\dr$.
As a set of chronicles, it is obtained by collecting all the results of execution:

$$\begin{array}{lll}
\dl\phi,\Psi\dr & = & \setc{q(+,\xi,I)}{\xst{\Psi',\Psi''}{\coupe{\phi}{\Psi}_\epsilon\longrightarrow^\star
 \coupe{(+,\xi,I)\cdot\Psi'}{\Psi''}_q}}\\
&& \union\;
\setc{q\,\Omega}{\coupe{\phi}{\Psi}_\epsilon\longrightarrow^\star \:!\,q\,\Omega}\\
&& \union\;
\setc{q\,\demon}{\coupe{\phi}{\Psi}_\epsilon\longrightarrow^\star \: !\,q\,\demon}\\
&& \union\;
\setc{q\,\Omega}{\mbox{computation does not terminate and }q\mbox{ is the maximal index reached}}
\end{array}$$

Note that we have added a second case of divergence: the divergence due to the cases where the machine does not terminate. To be more precise, we are speaking of non-termination due to rule (R) only (at index $q$), not of non-termination due to a potentially infinite branch of the normal form (recall that we are in a demand-driven setting).

Therefore, there are two sorts of $\Omega$ in the normal form: those that are \guil{created} (by the magic of a mathematical formula, not by some computation observed in a finite time), and those that come from the $\Omega's$ explicitly present in $\phi$ or $\Psi$.
The overloading of the two situations is really what makes the difference between $\Omega$ and $\demon$ that otherwise behave in a dual way (cf. e.g. the definition of $\leq^L$ and $\leq^R$). It is also the reason for the name ${\it Fid}$ given by Girard to the partial design $\Omega$: you must have faith ({\em Fides} in Latin) to wait for a result that might never come.

\smallskip
We still have to establish that $\dl\phi,\Psi\dr$ is a design, not only a pseudo-design, and hence we should exhibit a typing proof of $\dl\phi,\Psi\dr$. We first have to type the states of the machine, which we can also call nets, using Girard's terminology inspired from proof nets. We have seen that a state, or net, consists of a positive design $\phi$ and a collection $\Psi'$ of negative designs. We stick to this situation in the following definition.

\begin{definition} \label{designs-net}
A (positive) net is given by a positive design $\phi:(\vdash\Lambda)$, called the principal  design of the net, and a finite set $\Psi$ of negative designs such that:
\begin{enumerate} 
\item the addresses appearing  in the bases of $\phi$ and of the elements of $\Psi$ are pairwise  disjoint or equal;
\item an address may appear at most twice in these bases;
\item if an address appears twice, then it appears once on the left of $\:\vdash$ and once on the right of $\:\vdash$, thus forming a {\em cut};
\item The graph whose set of vertices is $\set{\phi}\union\Psi$ (or the associated set of bases) and whose edges are the cuts is acyclic.
\end{enumerate}
We denote the net by $\reseau{\phi}{\Psi}$.
\end{definition}

Girard requires moreover connectivity. We do not make this requirement explicit here, but it is clear that what interests us is the connected component of 
$\phi$. In this connected component, all the $\xi$'s on the left of $\:\vdash$ are cut.
This is easily shown by induction on the length of a path starting at $\vdash \Lambda$.
The only interesting case is when the path has reached a node of the form $\xi'\vdash\Lambda'$ and proceeds from there to a node of the form $\xi''\vdash\Lambda''$ with a cut on $\xi'$. But then by induction and by condition 2, the latter node must have been visited by the path earlier and must be the one that witnesses  that $\xi'$ is cut. Therefore, we know by induction that $\xi''$ is cut.

Hence all the non-cut addresses of the designs of the connected component stand to the right of  $\vdash$, in $\phi$ or in one of the elements of $\Psi$. Let $\Lambda$ be the set of these addresses. We set $\reseau{\phi}{\Psi}:(\vdash\Lambda)$. (It is thus the connected component that we actually type, but it is more convenient not to throw away explicitly the elements of the other connected components.) 

\begin{proposition}
If $\reseau{\phi}{\Psi}$ is a net
and if $\reseau{\phi}{\Psi}:(\vdash\Lambda)$, then $\dl\phi,\Psi\dr:(\vdash\Lambda)$.
\end{proposition}

\Proof We only prove that  rule (R) preserves typing, which  is enough to prove the statement in the case of a net of type $\vdash$ (see Remark \ref{var-tete-bien-typee}). Suppose that $\coupe{(+,\xi,I)\cdot \Psi'}{\Psi}:(\vdash\Lambda)$. We have to prove
$\coupe{\Psi_{\xi,I}}{\Psi'\union(\Psi\setminus\xi)}:(\vdash\Lambda)$.

\Proofitem{\mbox{Condition 4.}}
We have $(+,\xi,I)\cdot\Psi':(\vdash\xi,\ldots)$, and $\Psi:(\ldots,(\xi\vdash\ldots),\ldots)$, hence
$\Psi':\setc{(\xi i\vdash\ldots)}{i\in I}$  and
$(\Psi)_{\xi,I}:(\vdash\xi\star I,\ldots)$. Thus we remove the bases $(\vdash\xi,\ldots)$ and $(\xi\vdash\ldots)$ from the graph and replace them by the base $(\vdash\xi\star I,\ldots)$ and the bases
$(\xi i\vdash\ldots)$. The edge that connected $(\vdash\xi,\ldots)$ to $(\xi\vdash\ldots)$ is replaced by edges that connect 
$(\vdash\xi\star I,\ldots)$ to each of the $(\xi i\vdash\ldots)$'s. Note that these new edges form a connnected configuration.
The edges that once arrived to $(\xi\vdash\ldots)$ now arrive to 
$(\vdash\xi\star I,\ldots)$, and those that arrived to $(\vdash\xi,\ldots)$ are now either disconnected or arrive to one of the $(\xi i\vdash\ldots)$'s.
If there was a cycle in the new graph, one would then easily construct one in the former graph (to which we can go back by amalgamating the bases $(\xi i\vdash\ldots)$).

\Proofitem{\mbox{Conditions 2 and 3.}} The two occurrences of $\xi$ disappear, and are replaced by pairs of occurrences of each $\xi i$. \qed

\begin{remark} \label{var-tete-bien-typee}
A consequence of the previous proposition is that when the weak reduction machine starting from $\reseau{\phi}{\Psi}:(\vdash\Lambda)$ stops in a state $\coupe{(+,\xi_n,I_n)\cdot \Psi'_n}{\Psi_n}$, where $\Psi_n$ does not accept $\xi_n$, then $\xi_n\in\Lambda$. Indeed,
by type preservation we have  $\coupe{(+,\xi_n,I_n)\cdot \Psi'_n}{\Psi_n}:(\vdash\Lambda)$.
On the other hand we have  $(+,\xi_n,I_n)\cdot \Psi_n:(\vdash\Lambda_0)$ and 
$\Psi'_n:\set{\zeta_1\vdash\Lambda_1,\ldots,\zeta_m\vdash\Lambda_m}$ with $\xi_n\in \Lambda_0$ and
$\xi_n\nin\set{\zeta_1,\ldots,\zeta_m}$, that is, $\xi_n$ is not cut. Hence $\xi_n\in\Lambda$.
 In particular, when $\Lambda$ is empty, then the weak reduction machine can only end with an $\Omega$ or a $\demon$, and there is no need for strong reduction -- this situation is quite similar to the situation in a functional programming language like CAML, whose implementation is based on weak reduction, which is complete for evaluating observable results of basic types. In the sequel, we shall often place ourselves in such a situation, with $\phi: (\vdash\xi)$, $\Psi=\set{\psi}$, and $\psi:(\xi\vdash)$, for some $\xi$.
\end{remark}





\begin{definition}
For $\phi:(\vdash \xi)$ and $\psi:(\xi\vdash)$,
we write $\phi\bot\psi$ (or $\psi\bot\phi$) if $\dl\phi,\psi\dr=\demon$
and we say that $\phi$ is {\em orthogonal} to $\psi$.  For a set  $A$ of designs on the same base $\vdash \xi$ (resp. $\xi\vdash$), we write
$A^\bot=\setc{\psi}{\qqs{\phi\in A}{\phi\bot\psi}}$ (resp. $A^\bot=\setc{\phi}{\qqs{\psi\in A}{\phi\bot\psi}}$), and we write simply  $\phi^\bot$ for $\set{\phi}^\bot$.
\end{definition}

The definition of orthogonality is not limited to bases of the form $\vdash\xi$ and $\xi\vdash$:
more generally, if $\phi:(\vdash \xi_1,\ldots,\xi_n)$, 
$\psi_1:(\xi_1\vdash)$,\ldots, $\psi_n:(\xi_n\vdash)$, and $\dl\phi,\set{\psi_1,\ldots,\psi_n}\dr=\demon$, then we say that  $\phi$ is orthogonal to $\set{\psi_1,\ldots,\psi_n}$. 

\begin{exercise} \label{skunk-daimon}
Show that $\demon$ is the only design which is orthogonal to the skunk 
(cf. Definition \ref{dai-skunk-ram-dir}).
\end{exercise}

\begin{exercise} \label{norm-weak}
Show that if $I\inc J$, $\phi:(\vdash I)$ and $\psi_j:(j\vdash)$ for all $j\in J$,  then $\dl\phi,\setc{\psi_i}{i\in I}\dr=\dl\phi,\setc{\psi_j}{j\in J}\dr$
(taking  $\phi$ to be of type $\vdash J$ on the right side of this equality).
\end{exercise}

\begin{exercise}
Rule (R) does not make explicit the alternation between (the subdesigns of) $\phi$ and $\psi$: the positive actions that guide computation are indeed alternating. The following variant of rule (R) (to be applied systematically after the second step) makes this alternation explicit:

$$
\seql{(R')}{\coupe{(+,\xi,I)\cdot \Psi'}{\Psi}\longrightarrow \coupe{(+,\xi_1,I_1)\cdot \Psi'_1}{\Psi_1}\quad\quad (\Psi_1 \mbox{ accepts }\xi_1)}{\coupe{(+,\xi_1,I_1)\cdot \Psi'_1}{\Psi_1}\longrightarrow \coupe{(\Psi_1)_{\xi_1,I_1}}{\Psi'_1\union(\Psi\setminus\xi)}}
$$

\noindent
(1)Using this rule, check that the execution of the above example unrolls now as follows:

$$\begin{array}{lll}
\coupe{\phi}{\psi} & \longrightarrow & \coupe{\phi_1}{\Psi'}\\
& \longrightarrow & \coupe{\phi_2}{\Psi_1}\\
& \longrightarrow &  \coupe{\phi_3}{(\Psi'\setminus\xi i_1)\union\Psi_2}
\end{array}$$

\noindent (2) Show formally that  the machine obtained with (R') is equivalent to the one defined with (R).
\end{exercise}

\begin{exercise}  What about normalizing a negative design against a negative design? (Hint: Use strong reduction.)
\end{exercise}

\medskip
\Proofitem{\mbox{\bf Normalization in concrete syntax.}}
How does normalization look like, when described in terms of the concrete rather than the abstract syntax? We let the reader convince himself that the affine (weak reduction) machine given above gets mapped through the bijective correspondence between the concrete syntax and the abstract syntax to the following one:

$$\coupe{(x\cdot I)\setc{M_i}{i\in I}}{\rho\union(x\leftarrow\set{\ldots,I=\lbd\setc{x_i}{i\in I}.P_I,\ldots}}
\longrightarrow \coupe{P_I}{\rho\union(\Union_{i\in I}(x_i\leftarrow M_i)}$$

\noindent
In this rule, $\rho$ stands for an environment which is a function from a finite set of variables to terms, described as a set of bindings of the form $(x\leftarrow M$), and the union symbols stand for disjoint unions.  The initial state corresponding to $\coupe{\phi}{\psi}$ is here of the form $\coupe{P}{\set{(x\leftarrow M)}}$,
for some $P:(\vdash x:\xi,\Lambda_1)$ and $M:(\xi\vdash\Lambda_2)$.

\smallskip
The crucial use of the affine restriction lies in the fact that $x$ does not occur in any of the $M_i$'s, and hence that the binding for $x$ is consumed after the application of the rule.
But the general case of  non necessarily affine pseudo-designs in concrete syntax is not much more difficult to describe, by means of a very similar abstract machine, which is an instance of the (stack-free) Krivine machine described in 
\cite{CH98}:

$$\coupe{(x\cdot I)\setc{M_i}{i\in I}}{\rho}
\longrightarrow \coupe{P_I}{\rho_I\union(\Union_{i\in I}(x_i\leftarrow \coupe{M_i}{\rho})}$$

\noindent 
where $\rho(x)=\set{\ldots,I=\coupe{\lbd\setc{x_i}{i\in I}.P_I}{\rho_I},\ldots}$.

\smallskip
The main change with respect to the affine machine is that now an environment maps variables to closures, which are used in the implementation of functional programming languages to keep track of a code together with the environment for its free variables. Note that we keep now all the environment to be available for the $M_i$'s, even the binding for $x$, as $x$ may appear in any of the $M_i$'s.

\begin{exercise} \label{fax-renaming}
Show that, for any positive  $P$ of appropriate type, the (strong version of) the machine reduces
$\coupe{P}{\set{x'\leftarrow {\it Fax}_{\xi,x:\xi'}}}$ to $P[x'\leftarrow x]$  ($\alpha$-renaming).
\end{exercise}

\medskip
\Proofitem{\mbox{\bf Token machine: normalization as a visit.}}
We now give our third version of normalization, which is more intuitive. It is described in terms of pushing a token through 
the designs (see also \cite{Faggian02}). Under this interpretation, the net formed of $\phi$ and $\psi$ remains fixed, and information 
(in the form of a mark or token placed on a single node) flows through it. Initially, the token is placed at the root of $\phi$. In order to identify the nodes of $\phi$ and $\psi$, we introduce the following notation for occurrrences of actions:

$$\begin{array}{ccc} 
(\Omega)_\epsilon=\Omega && (\demon)_\epsilon=\demon
\\\\
((+,\xi,I)\cdot\setc{\psi_{\xi i}}{i\in I})_\epsilon=(+,\xi,I) && ((+,\xi,I)\cdot\setc{\psi_{\xi i}}{i\in I})_{iu}=(\psi_{\xi i})_u
\\\\
(\setc{(-,\zeta,J)\phi_J}{J\in{\cal P}_f(\omega)})_J=(-,\zeta,J) && (\setc{(-,\zeta,J)\phi_J}{J\in{\cal P}_f(\omega)})_{J1u}=(\phi_J)_u
\end{array}$$

\noindent
Thus an occurrence is a word over an alphabet whose letters are either $i$, $I$, or 1, and which 
appear in the order  $i_1I_11i_2I_21i_3\ldots$.  
The token is formalized as an occurrence $u$ of either $\phi$ or $\psi$, which we write
$(L,u)$ (resp. $(R,u)$) if the occurrence is in $\phi$ (resp. $\psi$).
The token machine maintains a set of pairs $((L,u),(R,v))$ where the actions $(\phi)_u$ and $(\psi)_v$ are opposite, and where each $u,v$ occurs at most once. These pairs are called {\em bindings} (in a sequent calculus description of cut-elimination (cf. part I, section 3), they represent the successive cuts).
Initially, the token is at occurrence $(L,\epsilon)$, and the set of bindings is empty.
The token game follows the tree structure and the pointer structure of $\phi$ and $\psi$, and the bindings.
The rules are as follows:
\begin{itemize}
\item from $(L,\epsilon)$, with $(\phi)_{\epsilon}=(+,\xi,I)$, move to $(R,I)$ and place
$(((L,\epsilon),(R,I))$ in the list of bindings;
\item from $(R,u)$ such that $(\psi)_u$ is a negative action, move to $(R,u1)$;
\item from $(R,u)$ such that $(\psi)_u=(+,\zeta i,I)$, $u$ points to $v$, and $((L,v'),(R,v))$ is in the set of bindings, then move to $(L,v'\,i\,I)$ and add $((L,v'\,i\,I),(R,u))$ to the set of bindings;
\item  from $(L,u)$ such that $(\phi)_u=(+,\zeta i,I)$, $u$ points to $v$, and $((L,v),(R,v'))$ is in the set of bindings, then move to $(R,v'\,i\,I)$ and add $((L,u),(R,v'\,i\,I))$ to the set of bindings;
\item from $(L,u)$ such that $(\phi)_u$ is a negative action, move to $(L,u1)$.
\end{itemize}

\noindent For example, the execution of our running example goes as follows, with the corresponding actions (recall that $\zeta=\xi i_1 i$):

$$\begin{array}{ccccccccc}
(L,\epsilon) & (R,I) & (R,I1) & (L,i_1I_1) & (L,i_1I_11) & (R,I1iJ) &  (R,I1iJ1) & (L,i_2\,I_2) & \ldots\\\\
(\phi)_\epsilon &  (\psi)_I & \psi_{I1} & (\phi)_{i_1I_1} & (\phi)_{i_1I_11} & (\psi)_{I1iJ} & (\psi)_{I1iJ1} & (\phi)_{i_2I_2}& \ldots\\
= & = & = & = & = & = & = & = & \ldots\\
(+,\xi,I) & (-,\xi,I) & (+,\xi i_1,I_1) & (-,\xi i_1,I_1) & (+,\zeta,J) & (-,\zeta,J) & (+,\xi i_2,I_2) & (-,\xi i_2,I_2)& \;\ldots
\end{array}$$

The actions {\em visited} during normalization are all the actions $(\phi)_u$  such that the token reaches position $(L,u)$ and all the actions $(\psi)_u$ such that the token reaches position $(R,u)$. They determine two designs $\phi_1\inc\phi$ and $\psi_1\inc\psi$ (the {\em pull-back}, see Theorem \ref{stability-theorem}) which form a {\em balanced} pair  (i.e., their sets of actions are dual).
Remarkably, there is a converse to this (see Proposition \ref{balanced-visit}).

\begin{exercise}
Show that the first and third versions are equivalent, i.e., that they yield the same result of normalization for every pair of designs $\phi:(\vdash\xi)$ and $\psi:(\xi\vdash)$. (Hint: At some point, Proposition
\ref{pull-back-slice} has to be used.)
\end{exercise}

\bigskip
\Proofitem{\mbox{\bf Analytical theorems.}}
These are the following theorems:

\begin{enumerate} 
\item the  {\em associativity theorem}, that corresponds to Church-Rosser property;
\item the {\em separation theorem}, that corresponds to  (an  affine  version of)  B\"ohm's theorem;
\item the {\em monotonicity theorem}, that corresponds to  the syntactic continuity theorem in the $\lambda$-calculus, due to Welch and L\'evy, which states that B\"ohm trees commute with contexts
(see e.g. \cite[section 2.3]{AmaCur});
\item the {\em stability theorem}, that corresponds  to the syntactic stability theorem of the $\lambda$-calculus, due to Berry, which states that the B\"ohm tree function from partial terms to partial B\"ohm trees is stable 
(it enjoys actually the stronger property of being sequential, see e.g.  \cite[section 2.3]{AmaCur}).
\end{enumerate}

\begin{remark}
The tradition behind the third and the fourth theorems is more widely known through their
outsprings in denotational semantics \cite{AmaCur}: Scott 's continuous semantics, Berry 's stable  semantics. Berry's semantics was clearly motivated by the stability theorem, while the syntactic continuity theorem seems rather to have given an additional a posteriori confirmation   of the pertinence of Scott's theory, which was initially suggested by results in  recursion theory (theorems of Rice and of Myhill-Shepherdson).
\end{remark}

We start with the stability theorem.

\begin{theorem}[Stability] \label{stability-theorem}
If $r$ is a chronicle of $\dl\phi,\psi\dr$, then there exist $\phi_0\inc\phi,\psi_0\inc\psi$ minimum, called the {\em pull-back} of $r$ along normalization, such that $r\in \dl\phi_0,\psi_0\dr$.
\end{theorem}

\Proof We mark all the nodes visited during the (deterministic) normalization in order to obtain $r$: all these nodes must be present, otherwise normalization would diverge, moreover they suffice (just adding $\Omega$'s above non visited negative actions following visited positive actions, in order to obtain
pseudo-designs). One also has to check that these pseudo-designs are actually designs (omitted).
\qed

\medskip
It is well-known  that stability described in terms of pull-backs as above entails stability in algebraic terms: if $\phi_1,\phi_2\inc\phi$ and $\psi_1,\psi_2\inc\psi$, then
$$\dl\phi_1\inter\phi_2,\psi_1\inter\psi_2\dr=\dl\phi_1,\psi_1\dr\inter
\dl\phi_2,\psi_2\dr$$
(set intersection of the designs as sets of chronicles).  
 In particular, for 
$\phi_1,\phi_2\inc\phi$ based on, say, $\vdash \xi$, and $\psi$ based on $\xi\vdash$, we have:

$$(\phi_1\inter\phi_2)\bot\psi \quad\Leftrightarrow\quad \phi_1\bot\psi\;\;\mbox{and}\;\;\phi_2\bot\psi\;.$$

\noindent
The same properties are true for any bounded intersection of  $\phi_k$'s, $\psi_k$'s.
\begin{theorem}[Separation]
The following equivalence holds, for all designs  $\phi_1,\phi_2$ on the same base: $\phi_1\sqinc\phi_2$ if and only if $\phi_1^\bot\inc\phi_2^\bot$.
\end{theorem}

\Proof Let $\phi_1\sqinc\phi_2$. By transitivity, we may restrict our attention to  $\leq^R$ and to $\leq^L$. Let $\psi$ be such that
$\phi_1\bot\psi$. Let $\phi$ be the part of $\phi_1$ that is effectively visited during normalization (cf. the stability theorem). If $\phi_1\leq^R\phi_2$, $\phi$ is also included in $\phi_2$, and we thus also have  $\phi_2\bot\psi$. If $\phi_1\leq^L\phi_2$, and if $\phi$ is not included in $\phi_2$, then some subtree of $\phi$ has been replaced by $\demon$, but then the normalization will meet this $\demon$, i.e., normalization is more rapid with $\phi_2$ than with $\phi_1$. 

Reciprocally, suppose that  $\phi_1^\bot\inc\phi_2^\bot$, and that there exists a pair of chronicles,
one in $\phi_1$, the other in $\phi_2$, that are not prefix of each other. Their intersection  $q$ ends with a negative  action, since after a  positive action, the set of negative actions depends only on the previous positive action, not on the  design. 
By Exercise \ref{orders-chronicles}, it suffices to check that the only possible configurations are the following:  either the branch that continues in $\phi_1$ is reduced to $\Omega$, or  the branch that continues in $\phi_2$ is reduced  to $\demon$. 
We use the following construction, that allows us to explore a chronicle interactively.
We first define the (dual)  {\em view } of a  chronicle (supposing that the design is based on
$\vdash\epsilon$, for simplicity):
 
$$\begin{array}{l}
{\it view}(r(-,\zeta,J)) = {\it view}(r)(+,\zeta,J)\\
{\it view}((+,\epsilon,I)) = (-,\epsilon,I)\\
{\it view}(r(-,\xi,J)\cdots(+\xi j,I)) = {\it view}(r)(+,\xi,J)(-,\xi j,I))
\end{array}$$

\noindent
We next define the following designs (as sets of chronicles):
\begin{itemize}
\item 
${\it Opp}_{r}$ (for a positive chronicle $r$) consists of  the set of the views of the prefixes of $r$, plus ${\it view}(r)\,\demon$; we have that ${\it Opp}_{r}$ is the minimum design such that $r\bot{\it Opp}_{r}$, by construction;
\item ${\it Opp}_{q}$ (for a negative chronicle) consists of the set of the views of the prefixes of $q$, plus all the chronicles of the form
${\it view}(q)(-,\zeta,J)\Omega$; we have that ${\it Opp}_{q}$ is the minimum design such that $(q\,\demon) \bot{\it Opp}_{q}$, by construction.
\end{itemize}

\Proofitem{\mbox{\sc Back to the proof.}} Let $q$ be the maximum common prefix.
We distinguish three cases:
\begin{enumerate}
\item $q\,\demon\in\phi_1$. Then we have $(q\,\demon)\bot {\it Opp}_q$, hence $\phi_1\bot {\it Opp}_q$. We should then have $\phi_2\bot {\it Opp}_q$,
but the interaction between $\phi_2$ and ${\it Opp}_q$ follows $q$, and then has not ``enough fuel" to continue  in $\phi_2$. This case is thus impossible.
\item $r=q(+,\xi,I)\in\phi_1$. We reason similarly. We have $r\bot {\it Opp}_r$, and the interaction has not ``enough fuel" to continue  in $\phi_2$, except if $q\,\demon\in\phi_2$ (second allowed configuration). 
\item $q\,\Omega\in\phi_1$. This is the first allowed configuration. \qed\\
\end{enumerate}

\begin{remark}
We have made a crucial use of the affinity condition in the proof of the separation theorem. Consider a chronicle of the form
$q\:(+,\xi)\:(-,\xi i_1)\:\ldots\:(+,\xi)\:(-,\xi i_2)\!\ldots$,
forgetting ramifications for simplicity. Then we collect two views
$${\it view}(q)\: (-,\xi)\:(+,\xi i_1)\mbox{ and }{\it view}(q) \:(-,\xi)\:(+,\xi i_2)$$
that do not fit together in a design, since a strategy answers an Opponent's move uniquely. 
\end{remark}

\begin{theorem}[Associativity]
Let $\phi$, $\Psi_1$, and $\Psi_2$ be such that $\phi$ and $\Psi_1\union\Psi_2$ form a net.
Then we have:
$$\dl\dl\phi,\Psi_1\dr,\Psi_2\dr=\dl\phi,\Psi_1\union\Psi_2\dr\;.$$
\end{theorem}

\Proofhint Let $r$ be a chonicle of $\dl\phi,\Psi_1\union\Psi_2\dr$, with pullback $\phi',\Psi'_1,\Psi'_2$.
One shows that $r$ can be pulled back along normalization of $\coupe{\dl\phi,\Psi_1\dr}{\Psi_2}$, with a pullback $\phi',\Psi''_2$ such that $\Psi'_2=\Psi''_2$ and the pullback of $\phi'$ along normalization of
$\dl\phi,\Psi_1\dr$ is $\phi',\Psi'_1$. A similar statement can be proved in the other direction.
These statements are best proved using a strong reduction version of the token machine. Intuitively, the machine  execution corresponding to the left hand side does some extra work with respect to the execution on the right hand side net:  as the execution proceeds, it records the part of the normal form of the net $\coupe{\phi}{\Psi_1}$ that is being built. \qed

\begin{theorem}[Monotonicity]  If $\phi_1\sqinc\phi_2$ and
$\psi_{1,1}\sqinc \psi_{2,1},\ldots, \psi_{1,n}\sqinc \psi_{2,n}$, then 
$$\dl\phi_1,\set{\psi_{1,1},\ldots,\psi_{1,n}}\dr\sqinc\dl\phi_2,\set{\psi_{2,1},\ldots,\psi_{2,n}}\dr\;.$$
\end{theorem}

\Proof
We have to show that $\dl\phi_1,\set{\psi_{1,1},\ldots,\psi_{1,n}}\dr\bot\Psi$ implies $\dl\phi_2,\set{\psi_{2,1},\ldots,\psi_{2,n}}\dr\bot\Psi$, for all $\Psi$.
By associativity, this amounts to deduce
$\dl\phi_2,\set{\psi_{1,1},\ldots,\psi_{1,n}}\union\Psi\dr=\demon$ from $\dl\phi_1,\set{\psi_{2,1},\ldots,\psi_{2,n}}\union\Psi\dr=\demon$, which is a consequence of the assumptions (recall that any increase in the order $\sqinc$ only fastens normalization).  \qed

\medskip
We end the section by showing that normalization explores only multiplicative  parts.

\begin{proposition} \label{pull-back-slice}
The pull-back of a chronicle is always a slice (cf. Definition \ref{slice}). More generally, the pullback of a slice is a slice. Moreover, each node of this slice is visited only once.
\end{proposition}

\Proof We limit ourselves to a normalization between $\phi:(\vdash\xi)$ and $\psi: (\xi\vdash)$ (and hence to the pullback of $\demon$).
For two negative actions $(-,\xi i,I_1)$ and $(-,\xi i,I_2)$ with the  same focus and above the same  positive action $(+,\xi,I)$ to be visited,  
we need on the other side two positive actions $(+,\xi i,I_1)$ and $(+,\xi i,I_2)$ pointing to the same negative action $(-,\xi,I)$.  We reason by minimal  counter-example, where minimality is taken with respect to  the time at which the hypothetical second visit takes place.
Typing excludes that these two positive actions be one above another (the address is consumed after the first visit). Thus they must appear in incompatible positions in the tree, and the divergence starts from a positive action. 
Then typing implies that the first two negative actions on the two diverging paths must have the same focus (disjointness of contexts). These two actions must have been visited before, a contradiction to minimality.  Note that the reasoning applies a fortiori with $I_1=I_2$, which proves the last part of the statement.
\qed

\medskip
Notice that Proposition \ref{pull-back-slice} implies that the token machine terminates on finite designs (a property which is obvious with our first formalization of normalization).

\begin{proposition} \label{balanced-visit}
 If two finite slices $\phi$ and $\psi$ are such that their underlying set of actions are opposite, then the normalization of $\coupe{\phi}{\psi}$ visits all of $\phi$ and $\psi$.
\end{proposition}

\Proof (A more general statement is proved in \cite[Proposition 1]{LS00}). Note that
since we deal with slices, we can omit the ramifications, and name the actions simply with their sign and their focus. We say that two slices satisfying the condition in the statement form a {\em balanced pair}:
$(+,\xi)$ (resp. $(-,\xi)$) occurs in $\phi$ if and only if $(-,\xi)$ (resp. $(+,\xi)$) occurs in $\psi$.

The proof is by contradiction. 
If the normalization stops before having explored $\phi$ and $\psi$ entirely, it stops  having visited $\phi'\subseteq \phi$ and $\psi'\subseteq \psi$, where $\phi'$ and $\psi'$ form a balanced pair, and we have, say, $\phi'\neq \phi$, but then also $\psi'\neq \psi$ since $\phi'$ and $\psi'$ have the same addresses. Consider the actions on the border, i.e., which are not in $\phi'$ nor in $\psi '$, but whose father is in $\phi'$ or $\psi'$: they must be negative, since otherwise the normalization procedure would have visited them. So let us pick an action  $(-,\xi_1)$ on the border, say in $\phi\setminus\phi'$. Then also $(+,\xi_1)$ occurs somewhere in $\psi\setminus \psi'$. Consider the chronicle (determined by) $(+,\xi_1)$, and let $(-,\xi_2)$ be the action of this chronicle which lies on the border. We can continue this 
and build an infinite sequence of actions  $(-,\xi_n)$.  Now, for all $n$, we have $\xi_n=\xi'_n i_n$  and $(+,\xi'_n)\in \phi'\union\psi'$, for some $\xi'_n$ and $i_n$ (focalization condition, cf. Remark \ref{design-chronicles}). We also have that $(-,\xi'_n)\in  \phi'\union\psi'$, since $\phi'$ and $\psi'$ form a balanced pair.
The action  $(-,\xi'_n)$ appears on the chronicle $(+,\xi_n)$
(subaddress condition), and hence by construction $(-,\xi'_n)$ occurs before $(+,\xi'_{n+1})$ on this chronicle.
It follows that $(-,\xi'_n)$  is visited before $(+,\xi'_{n+1})$ during normalization.
Since $(+,\xi'_n)$ is visited right before $(-,\xi'_n)$, we have that
$(+,\xi'_{n+1})$ is visited strictly after $(+,\xi'_{n})$.  But this contradicts the termination of the normalization procedure, which as we have seen is a consequence of Proposition \ref{pull-back-slice}. \qed

\medskip 
Several characterizations of those designs $\phi$ which can be entirely visited during normalization against a counter-design $\psi$ are given in \cite{Faggian04}.

\begin{exercise} \label{fermeture}
Show that in presence  of the separation property, associativity is equivalent to (a general formulation of) the following statement (called {\em closure principle}). If $\phi:(\vdash\xi)$ and 
$\psi:(\xi\vdash\lambda)$, then $\dl\phi,\psi\dr$ is the unique design $\phi'$ such that $\dl\phi',\psi'\dr=\dl\phi,\set{\psi,\psi'}\dr$
for all $\psi':(\lambda\vdash)$.
\end{exercise}

\section{Behaviours} \label{behaviours}

A  positive {\em behaviour} on a base $\vdash\Lambda$ (resp. a negative behaviour on a base
$\xi\vdash\Lambda$)  is a set  ${\bf G}$ of designs of type $(\vdash\Lambda)$ (resp. ($\xi\vdash\Lambda$)) closed by taking the bi-orthogonal, i.e., ${\bf G}={\bf G}^{\bot\bot}$. In most cases, we suppose that the base is of the form  $\vdash\xi$ or $\xi\vdash$. 
Equivalently, a behaviour, say, on the base $\vdash\xi$ (resp. ($\xi\vdash$)) is a set of the form $A^\bot$, where $A$ is an arbitrary set of designs of type ($\xi\vdash$) (resp. ($\vdash\xi$)).
Throughout the section, we take $\xi=\epsilon$.
Here are some examples of behaviours:
\begin{itemize}
\item If $A=\emptyset$, then $A^\bot$ is the set of all the designs of type ($\epsilon\vdash$) (resp. ($\vdash\epsilon$)): this behaviour is denoted $\mbox{\boldmath $\top$}$.
\item If $A=\mbox{\boldmath $\top$}$ is negative (resp. positive), it is easy to see that $A^\bot=\set{\it Dai}$
(resp. $A^\bot=\set{{\it Dai}^-}$). 
 Hence a behaviour is never empty, it always contains ${\it Dai}$ or ${\it Dai}^-$.
\item If $\phi$ is a design, the smallest behaviour that contains it is:
$$\set{\phi}^{\bot\bot}=\setc{\phi'}{\phi\sqinc\phi'}\;.$$ 
Indeed, for any $A$, we have $A^\bot=\setc{\phi'}{A\inc \phi'^\bot}$, so in particular $\set{\phi}^{\bot\bot}=\setc{\phi'}{\phi^\bot\inc \phi'^\bot}=\setc{\phi'}{\phi\sqinc\phi'}$.
\end{itemize}

\medskip\noindent
Below, we list a few closure properties of behaviours, which are easily checked:
\begin{itemize}
\item If ${\bf G}$ is a behaviour and if $\phi\in{\bf G}$ and $\phi\sqinc \phi'$, then $\phi'\in{\bf G}$ (by the separation theorem).
\item If ${\bf G}$ is a behaviour and if $\phi_k$ is a bounded family of designs (considered as sets of chronicles) of ${\bf G}$, i.e., $\xst{\phi}{(\qqs{k}{\phi_k\inc\phi})}$), then their intersection is a design of ${\bf G}$ (by the stability theorem).
\item Every intersection of behaviours is a behaviour (as, say,  $A^\bot\inter B^\bot=(A\union B)^\bot$).
\end{itemize}

\begin{definition}[incarnation]
If ${\bf G}$ is a behaviour and if $\phi\in{\bf G}$, we call the incarnation  of $\phi$ in $\bf G$ the smallest design $\inc\phi$ of ${\bf G}$ (whose existence follows from the second closure property listed above). We denote it as
$|\phi|_{\bf G}$ or simply $|\phi|$. An incarnated design is a design $\phi$ such that $\phi=|\phi|$. We set 
$|{\bf G}|=\setc{\phi\in{\bf G}}{\phi=|\phi|_{\bf G}}$.
\end{definition}

The stability theorem and the definition of behaviour allow us to give a more operational characterization of the incarnation: $|\phi|_{\bf G}$ is the set of chronicles that are visited during a normalization of $\phi$ against some design $\psi$ of ${\bf G}^\bot$ (different designs may be used for witnessing the visit of different chronicles).

Incarnation is clearly contravariant: if ${\bf G}\inc{\bf H}$ and $\phi\in{\bf G}$, then $|\phi|_{\bf H}\inc|\phi|_{\bf G}$.
The extreme case is $|\psi|_{\mbox{\boldmath $\top$}}={\it Skunk}$ (negative  base).

\begin{lemma} \label{incarnation-orthogonal}
For any behaviour, we have ${\bf G}^\bot=|{\bf G}|^\bot$.
\end{lemma}
\Proof We get ${\bf G}^\bot\inc|{\bf G}|^\bot$ by contravariance. Now let $\psi\in|{\bf G}|^\bot$ and let $\phi\in G$. Then $\psi\bot |\phi|_{\bf G}$, and hence $\psi\bot\phi_G$ by monotonicity. \qed

\medskip
\Proofitem{\mbox{\bf Additives.}} We now have enough material  to return to types and logic. 
The paradigm is that of behaviours as types. The rest of the section is devoted to constructions on behaviours, corresponding to those of linear logic, and more. Some of these constructions are
reminiscent of phase semantics (cf. part I, section 5), but the present framework is obviously much richer.

\begin{definition}[Intersection, Union] \label{inter-union}
If ${\bf G}_k$ is a family of behaviours on the same base, the set intersection of this familly is a behaviour (cf. above), that we call the intersection of these behaviours, notation $\Inter_k {\bf G}_k$. 
We define the union of behaviours $\bigsqcup_k {\bf G}_k$ as the bi-orthogonal of their set union: $\bigsqcup_k {\bf G}_k=(\Union_k {\bf G}_k)^{\bot\bot}$.
\end{definition}

The ordinary additives correspond to the particular case where the connectives are applied to disjoint behaviours -- a notion that we define now (using some of the designs of Definition  \ref{dai-skunk-ram-dir}).

\begin{definition}[Directory] \label{directory}
A directory is a set of ramifications (i.e., a subset of ${\cal P}_f(\omega)$). If ${\bf G}$ is a positive behaviour on $\vdash\epsilon$, we define the directory $\dirct{\bf G}$ as follows: $\dirct{{\bf G}}=\setc{I}{{\it Ram}_{(\epsilon,I)}\in{\bf G}}$. If ${\bf G}$ is negative, we define $\dirct{\bf G}$ by the equation ${\it Dir}_{\dirct{\bf G}}=|{\it Dai}^-|_{\bf G}$.
\end{definition}

The following  properties show the relevance of this definition:
\begin{itemize}
\item If ${\bf G}$ is positive, then $\dirct{\bf G}$ is the set of the $I$'s such that $\bf G$ contains a design beginning with $(+,\epsilon,I)$. Let us check this. If $\phi\in{\bf G}$ begins with $(+,\epsilon,I)$, then 
$\phi\sqinc{\it Ram}_{(\epsilon,I)}$, hence ${\it Ram}_{(\epsilon,I)}\in{\bf G}$, i.e., $I\in\dirct{\bf G}$. The reciprocal is immediate.
\item We  have $\dirct{{\bf G}^\bot}=\dirct{\bf G}$. The proof goes as follows. Since ${\bf G}^{\bot\bot}= {\bf G}$, we may suppose that $\bf G$ is positive. We have $I\in \dirct{{\bf G}^\bot}$ if and only if there exists $\phi\in {\bf G}$ such that the normalization of $\coupe{\phi}{{\it Dai}^-}$ explores the branch
$(-,\epsilon,I)$ of ${\it Dai}^-$, i.e.,  such that  $\phi$ begins with $(+,\epsilon,I)$. But by the previous property, this amounts to $I\in\dirct{\bf G}$.
\item  If {\bf G} is a negative behaviour and if $\psi$ is incarnated in {\bf G} on base $\xi\vdash$, then ${\it Dir}({\bf G})=\setc{I}{(-,\xi,I)\mbox{ is an initial action of }\psi}$.
Indeed, each $\phi$ of ${\bf G}^\bot$ induces a visit on $\psi$, which starts with
a $(-,\xi,I)$ matching $\phi$'s initial action: collecting all these visits together (which is the definition of incarnation), we get the statement.
\end{itemize}

\begin{definition}[Disjoint behaviours]
We say that two behaviours ${\bf G}$ and ${\bf G}'$ on the same base are 
disjoint if $\dirct{\bf G}$ and $\dirct{{\bf G}'}$ are disjoint sets.
\end{definition}


 

If two negative behaviours {\bf G} and {\bf H} are disjoint (that is,
${\it Dir}({\bf G})\inter{\it Dir}({\bf H})=\emptyset$), and if $\psi_1$ and $\psi_2$ are respective incarnated designs of {\bf G} and {\bf H}, their union is well-defined. Moreover, it is obviously a design of 
${\bf G}\inter{\bf H}$. This actually defines a bijection between $|{\bf G}|\times|{\bf H}|$ and
$|{\bf G}\inter{\bf H}|$. Hence in the disjoint case, intersection is product!  Girard calls this
striking property the ``mystery of incarnation".

\begin{lemma} \label{orthogonal-of-with}
If {\bf G} and {\bf H} are disjoint negative behaviours, then $({\bf G}\inter{\bf H})^\bot={\bf G}^\bot\union{\bf H}^\bot$.
\end{lemma}

\Proof By contravariance, we have ${\bf G}^\bot\inc({\bf G}\inter{\bf H})^\bot$ and 
${\bf H}^\bot\inc({\bf G}\inter{\bf H})^\bot$. It remains to show 
$({\bf G}\inter{\bf H})^\bot\inc{\bf G}^\bot\union{\bf H}^\bot$. Let $\phi\in ({\bf G}\inter{\bf H})^\bot$.
If $\phi=\demon$,  then $\phi$ belongs to any behaviour, and hence a fortiori to 
${\bf G}^\bot\union{\bf H}^\bot$. So we can suppose that $\phi$ starts with a positive action.
For any pair of incarnated designs $\psi_1\in{\bf G}$ and $\psi_2\in{\bf H}$, we have (cf.  above) $\psi_1\union\psi_2\in {\bf G}\inter{\bf H}$, and hence $\phi\bot(\psi_1\union\psi_2)$.
Then $\phi$'s initial action $(+,\xi,I)$ matches an initial action of (exclusively) either $\psi_1$ or $\psi_2$, say, not of $\psi_2$: then all normalizatiojn takes place in $\psi_1$, so we have in fact $\phi\bot\psi_1$. Now we can let $\psi_1$ vary over $|{\bf G}|$ while keeping $\psi_2$ fixed.  Then we have $\phi\bot\psi$ for all $\psi\in|{\bf G}|$, that is, $\phi\in |{\bf G}|^\bot$, and we conclude by Lemma \ref{incarnation-orthogonal}. \qed

\begin{definition}[Additive connectives]
If $\bf G$ and $\bf H$ are two disjoint negative (resp. positive) behaviours, we
rebaptize their intersection (resp. union) in the sense of definition
\ref{inter-union} as follows:
${\bf G}\inter{\bf H}={\bf G}\with{\bf H}$ (resp. ${\bf G}\bigsqcup{\bf H}={\bf G}\oplus{\bf H}$).
\end{definition}

\begin{proposition}
If ${\bf G}$ and ${\bf H}$ are negative and disjoint, then
$|{\bf G}\with {\bf H}| \approx |\bf G|\times |\bf H|$.
\end{proposition}

\Proof We have constructed above a mapping from $|{\bf G}|\times|{\bf H}|$ to ${\bf G}\inter {\bf H}$, which takes $\psi_1$ and $\psi_2$ and returns $\psi_1\union \psi_2$.  This design is incarnated, since the whole of $\psi_1$ is visited by normalisation against the designs of ${\bf G}^\bot$ (which are a fortiori designs of $({\bf G}\inter {\bf H})^\bot$), and similarly for $\psi_2$.  Hence we have defined a map from
$|{\bf G}|\times|{\bf H}|$ to $|{\bf G}\inter {\bf H}|$. This map is injective by the disjointness assumption.
We are left to show that it is surjective.
Let $\psi$ be an incarnated design of ${\bf G}\inter {\bf H}$, Then, by Lemma \ref{orthogonal-of-with}, we can write $\psi=\psi_1\union\psi_2$, where, say, $\psi_1=\setc{\psi_\phi}{\phi\in {\bf G}^\bot}$ (where $\psi_\phi$ is the part of $\psi$ visited during the normalization of
$\coupe{\phi}{\psi}$), which is incarnated in ${\bf G}$.  \qed

\begin{remark} Note that the cartesian product is associative only up to isomorphism, while intersection is isomorphic up to equality.  Girard points out that a way to get a better match is to redefine the product of two disjoint sets $X$ and $Y$ as follows:
$X\times Y =\setc{x\union y}{x\in X\mbox{ and }y\in Y}$.
Note that this definition makes also sense when the condition on $X$ and $Y$ is not satisfied, but
one then  does not obtain a product (in the category of sets) of $X$ and $Y$ anymore. 
\end{remark}

\medskip
The $\oplus$ connective has also a remarkable property: one can get rid of the bi-orthogonal. Moreover, it is  also a union at the level of incarnations (see Exercise \ref{plus-incarne-union}).
\begin{proposition}
Let {\bf G} and {\bf H} be positive and disjoint behaviours. Then ${\bf G}\oplus{\bf H}$ is simply the set union of ${\bf G}$ and ${\bf H}$.
\end{proposition}

\Proof This is an immediate consequence of Lemma \ref{orthogonal-of-with}:
$${\bf G}\union{\bf H}={\bf G}^{\bot\bot}\union{\bf H}^{\bot\bot}= ({\bf G}^\bot\inter{\bf H}^\bot)^\bot=
({\bf G}\union{\bf H})^{\bot\bot}\; .$$
\qed

\begin{remark} Girard calls {\em internal completeness} the situation when a new behaviour can be defined from other behaviours without using a bi-orthogonal. Indeed, given a set $A$ of designs, the set $A^\bot$ can be viewed as the set of the (counter-)models of $A$, and then $A$ being closed under bi-orthogonal means that everything valid was already there.
\end{remark}

What if two behaviours are not disjoint? Then we can still form a $\with$ or a $\oplus$, provided we force disjunction by copying the behaviours, or {\em delocating} them, as Girard says (see Exercises \ref{delocate} and \ref{delocated-with-plus}).

\smallskip
But intersections are not less useful than products. They have been introduced long ago in the study of models of untyped $\lambda$-calculi \cite{Coppo79,Krivine91}, and have been used to give semantic foundations to object-oriented  programming (see, e.g. \cite{FOO94}). We show here a simple example of how records can be encoded as designs, and how they can be observed by suitable behaviours.
Let us consider three fields: {\tt radius}, {\tt angle}, {\tt colour}. Assuming that we have fixed an origin and
two orthogonal vectors of equal norm, then giving a positive number $r$ as radius and a positive number $\phi\pmod{360}$ gives us a point in the plane, while giving an $r$ and a coulour, say, blue, gives us a blue circle. If the three fields are given a value, then we get a couloured point.
To encode these simple data, we shall use negative designs (recall that $\with$ is a negative connective), on base $\epsilon\vdash$. Let $I_1,I_2,I_3$ be arbirtrary ramifications (i.e., finite subsets of $\omega$), which are pairwise distinct, for example, $I_1=\set{i_1}, I_2=\set{i_2}, I_3=\set{i_3}$ with $i_1,i_2,i_3$ distinct. We shall use them to encode the three fields {\tt radius}, {\tt angle}, and {\tt colour}, respectively. 
Here is a red point on the horizontal axis, negative side, at distance 2 from the origin (yes, we encode here only denumerably many points..., and, say, red is encoded by 9):

$$\begin{array}{lll}
\psi & = &
\set{(-,\epsilon,\set{i_1})\,(+,i_1,\set{2})\cdot\set{{\it Skunk}_{i_12}},\\
&&\;(-,\epsilon,\set{i_2})\,(+,i_2,\set{180})\cdot\set{{\it Skunk}_{i_2\star(180)}},\\
&&\;(-,\epsilon,\set{i_3})\,(+,i_3,\set{9})\cdot\set{{\it Skunk}_{i_39}}}
\end{array}$$

\noindent 
Suppose that we are only interested in the underlying red circle (centered at the origin) -- yes, with this representation, the type of points is a subtype of the type of circles centered at the origin:  we simply forget the {\tt angle} component. Formally, we define the following behaviour ${\bf G}$ of coloured circles:

$${\bf G}\quad=\quad\set{(+,\epsilon,\set{i_1})\cdot\set{{\it Dai}^-_{i_1}}\;,\;
(+,\epsilon,\set{i_3})\cdot\set{{\it Dai}^-_{i_3}}}^\bot\;.$$

\noindent
We have $|\psi|_{{\bf G}} \;=\; \set{(-,\epsilon,\set{i_1})\,(+,i_1,\set{2})\cdot\set{{\it Skunk}_{i_12}}\;,
\;(-,\epsilon,\set{i_3})\,(+,i_3,\set{9})\cdot\set{{\it Skunk}_{i_39}}}$.
One could similarly define the behaviour ${\bf G'}$ of (uncoloured) points. Then ${\bf G}\inter{\bf G'}$ is the behaviour of coloured points, and this behaviour is not a product. We could also define the behaviours ${\bf G_1}$ and ${\bf G_2}$ of circles and colours, respectively, and then we recover ${\bf G}$ as the intersection of ${\bf G_1}$ and ${\bf G_2}$, which in this case is a product.

\bigskip
There is more material in \cite{LS00} and in \cite{Maur04}, to which we refer for further reading:
\begin{itemize}
\item In ludics, one can define different sorts of commutative and non-commutative tensor products of designs and of behaviours. The basic idea is to glue two positive designs $\phi_1$ and $\phi_2$ starting respectively with $(+,\xi,I)$ and $(+,\xi,J)$ into a design starting with $(+,\xi,I\union J)$, the problem being what to do on $I\inter J$: the non-commutative versions give priority to one of the designs, while the commutative versions place uniformly $\bot$  (resp. $\demon$) after all actions $(-,\xi k,L)$, where
$k\in I\inter J$.  The degenerate cases of the definition  (tensoring with ${\it Fid}$) are a bit tricky, and have led
Maurel to introduce an additional design, called negative divergence (see \cite{Maur04}).
\item Ludics suggests new kinds of quantifications, in a vein similar to what we have seen for additives.
\item Girard proves a full completeness result for a polarized version of MALL (cf. section \ref{ludics}) with quantifiers. To this aim, behaviours must be enriched with partial equivalence relations, to cope with the uniform, or parametric nature of proofs (think of axiom $\vdash X,X^\bot$ as parametric in the atom $X$).
\item Exponentials can be handled by introducing a quantitative framework, where positive actions
in designs are combined through formal probability trees: probabilities allow us to count repetitions, and to lift the separation property to a framework of designs with repetition and pointers (see \cite{Maur04}).
\end{itemize} 

\begin{exercise} \label{char-disjoint}
Show that two positive behaviours are disjoint if and only if their set intersection is $\set{\demon}$,
and that two negative behaviours ${\bf G}$ and ${\bf G}'$ are disjoint if and only if for all $\psi\in{\bf G}$ and $\psi'\in{\bf G}'$ we have $|\psi|_{\bf G}\inter|\psi'|_{{\bf G}'}=\emptyset$ (as sets of chronicles).
\end{exercise}

\begin{exercise} \label{plus-incarne-union}
Show that if {\bf G} and {\bf H} are positive and disjoint, then $|{\bf G}\oplus{\bf H}|=|{\bf G}|\union|{\bf H}|$.
\end{exercise}

\begin{exercise} \label{delocate}
What conditions should a function $\theta$ from addresses to addresses satisfy to be such that it induces a well defined transformation on designs, replacing every action, say, $(+,\xi,I)$ with $(+,\theta(\xi),J)$ such that $\theta(\xi)\star J=\setc{\theta(\xi i)}{i\in I}$? Such a function is called a {\em delocation} function.
\end{exercise}

\begin{exercise} \label{delocated-with-plus}
Show that for any two behaviours ${\bf G_1}$ and ${\bf G_2}$ on the same base, one can find two delocation functions $\theta_1$ and $\theta_2$ such that $\theta_1({\bf G_1})$ and 
$\theta_2({\bf G_2})$ are disjoint. 
\end{exercise}

\begin{exercise} \label{distrib-behaviours}
Show that as operations on behaviours $\otimes$ and $\oplus$ are such that $\otimes$ distributes over $\oplus$.
\end{exercise}

\subsection*{Acknowledgements}  I wish to thank Song Fangmin and the University of Nanjing for an invitation  in 
October 2001, which offered me the occasion to give a few lectures on linear logic,
and 
Aldo Ursini and the University of Siena for providing most hospitable conditions for me to write this survey article and to give the corresponding course.

\end{document}